\documentclass[12pt, draftclsnofoot, hidelinks, onecolumn]{IEEEtran}
\usepackage{epsfig,graphicx,subfigure,psfrag,amsmath,cases}
\usepackage{latexsym,amssymb,amsmath,epsfig,subfigure,algorithm,mathtools}
\usepackage{algorithmic}
\usepackage{color}
\usepackage{url}
\usepackage{scrtime}
\usepackage{cite}
\usepackage{hyperref}
\usepackage{subfigure}
\usepackage{bbding}
\usepackage{multicol}
\usepackage{bm}
\usepackage{xcolor}
\usepackage[framemethod=TikZ]{mdframed}
\usetikzlibrary{shadows}
\usepackage{environ}
\usepackage{varwidth}

\newlength{\MyMdframedWidthTweak}%
\NewEnviron{MyMdframed}[1][]{%
    \setlength{\MyMdframedWidthTweak}{\dimexpr%
        +\mdflength{innerleftmargin}
        +\mdflength{innerrightmargin}
        +\mdflength{leftmargin}
        +\mdflength{rightmargin}
        }%
    \savebox0{%
        \begin{varwidth}{\dimexpr\linewidth-\MyMdframedWidthTweak\relax}%
            \BODY
        \end{varwidth}%
    }%
    \begin{mdframed}[
        backgroundcolor=lightgray,
        shadow=true,
        shadowsize=4pt,
        roundcorner=5pt,
        userdefinedwidth=\dimexpr\wd0+\MyMdframedWidthTweak\relax,
        #1]
        \usebox0
    \end{mdframed}
}

\author{Zhiqiang Wei,~\IEEEmembership{Member,~IEEE,} Fan Liu,~\IEEEmembership{Member,~IEEE,} Chang Liu,~\IEEEmembership{Member,~IEEE,} Zai Yang,~\IEEEmembership{Senior Member,~IEEE,} Derrick Wing Kwan Ng,~\IEEEmembership{Fellow,~IEEE,} and Robert Schober, ~\IEEEmembership{Fellow,~IEEE} 
	\thanks{Z. Wei and Z. Yang are with the School of Mathematics and Statistics, Xi'an Jiaotong University, Xi'an 710049, China (e-mail: zhiqiang.wei@xjtu.edu.cn; yangzai@xjtu.edu.cn); F. Liu is with the Department of Electrical and Electronic Engineering, Southern University of Science and Technology, China (email: liuf6@sustech.edu.cn); C. Liu and D. W. K. Ng are with the School of Electrical Engineering and Telecommunications, the University of New South Wales, Australia (email: chang.liu19@unsw.edu.au; w.k.ng@unsw.edu.au); R. Schober is with the Institute for Digital Communications (IDC), Friedrich-Alexander University Erlangen-Nuremberg, Germany (email: robert.schober@fau.de). Part of the paper was presented at ICASSP 2022\cite{ZhiqiangICASSP}.}}

\title{Integrated Sensing, Navigation, and Communication for Secure UAV Networks with a Mobile Eavesdropper\vspace{-2mm}}

\newtheorem{T-Prob}{Transformed Problem}

\DeclareMathOperator{\Tr}{Tr}
\DeclareMathOperator{\Rank}{Rank}
\DeclareMathOperator{\maxo}{maximize}
\DeclareMathOperator{\mino}{minimize}
\DeclareMathOperator{\diag}{\mathrm{diag}}

\newcommand{\abs}[1]{\lvert#1\rvert}

\begin{document}
\maketitle

\vspace{-20mm}
\begin{abstract}
	\vspace{-4mm}
	This paper proposes an integrated sensing, navigation, and communication (ISNC) framework for safeguarding unmanned aerial vehicle (UAV)-enabled wireless networks against a mobile eavesdropping UAV (E-UAV).
	To cope with the mobility of the E-UAV, the proposed framework advocates the dual use of artificial noise transmitted by the information UAV (I-UAV) for simultaneous jamming and sensing to facilitate navigation and secure communication.
	In particular, the I-UAV communicates with legitimate downlink ground users, while avoiding potential information leakage by emitting jamming signals, and estimates the state of the E-UAV with an extended Kalman filter based on the backscattered jamming signals.
	Exploiting the estimated state of the E-UAV in the previous time slot, the I-UAV determines its flight planning strategy, predicts the wiretap channel, and designs its communication resource allocation policy for the next time slot.
	To circumvent the severe coupling between these three tasks, a divide-and-conquer approach is adopted.
	The online navigation design has the objective to minimize the distance between the I-UAV and a pre-defined destination point considering kinematic and geometric constraints.
	Subsequently, given the predicted wiretap channel, the robust resource allocation design is formulated as an optimization problem to achieve the optimal trade-off between sensing and communication in the next time slot, while taking into account the wiretap channel prediction error and the quality-of-service (QoS) requirements of secure communication.
	To account for the E-UAV state sensing uncertainty and the resulting wiretap channel prediction error, we employ a fully-connected neural network to model the complicated mapping between the state estimation error variance and an upper bound on the channel prediction error, which facilitates the development of a low-complexity suboptimal user scheduling and precoder design algorithm.
	Simulation results demonstrate the superior performance of the proposed design compared with baseline schemes and validate the benefits of integrating sensing and navigation into secure UAV communication systems.
	We reveal that the dual use of artificial noise can improve both sensing and jamming and that navigation is more important for improving the trade-off exploration between sensing and communications than communication resource allocation.\vspace{-2mm}
\end{abstract}

\begin{keywords}
	UAV, physical layer security, extended Kalman filter, resource allocation.\vspace{-4mm}
\end{keywords}

\section{Introduction}
\label{sec:intro}

Given the multitude of emerging aerial applications and the substantially reduced manufacturing costs, unmanned aerial vehicles (UAVs) have become a promising platform for wireless communications \cite{ZengAccessSky}.
It is expected that UAV-enabled wireless networks will play a key role in supplementing future cellular networks by providing ad-hoc/on-demand communication services to rural, disaster-affected, and hot-spot areas \cite{wu20205g}.
In contrast to conventional terrestrial networks with fixed infrastructure, UAV-enabled wireless networks \cite{ZhiqiangUAV} have the potential to seamlessly extend the existing network coverage thanks to their high maneuverability and high cruising speed.
Moreover, UAVs are more likely able to establish strong line-of-sight (LoS) channels to the ground terminals compared with their terrestrial counterparts\cite{ZengAccessSky}, which can be leveraged to improve communication performance.

Nevertheless, in practice, various technical challenges have to be carefully addressed to realize the full potential of UAV wireless networks.
First, although the controllable mobility of UAVs provides a new degrees of freedom to improve network performance, the navigation design is intricately coupled with the resource allocation design, which introduces a new challenge.
This issue has been studied extensively in the literature.
Assuming a simple free-space deterministic path loss model, the communication performance becomes predictable as a function of the UAV trajectory which enables an offline joint trajectory and resource allocation design.
For instance, in \cite{ZengUAV}, the authors employed a UAV to serve as a mobile relay node and optimized its trajectory and resource allocation policy to maximize the end-to-end system throughput.
Besides, the authors in \cite{ZengEEUAV} developed a power consumption model for fixed-wing UAVs and proposed a joint trajectory and resource allocation design to maximize the system energy efficiency.
Also, the authors in \cite{QingqingMultiUAV} investigated the case of multiple cooperating UAVs and jointly designed the user scheduling, UAV trajectory, and power allocation to maximize the minimum throughout among all ground terminals.
Furthermore, considering more practical stochastic channel models for the aerial-ground links, such as the probabilistic LoS channel model\cite{ZengAccessSky,HouraniProLoS} and the angle-dependent Rician fading channel model\cite{Changsheng3DRician}, an offline joint trajectory and resource allocation design is still possible by properly accounting for the inherent channel randomness.
In particular, the authors in \cite{ZengAccessSky} considered a probabilistic LoS channel model and adopted the average channel power gain for joint trajectory and resource allocation design.
Moreover, in \cite{Changsheng3DRician}, the authors assumed an angle-dependent Rician fading channel model and proposed to approximate the effective fading power via a data regression method to characterize the communication performance over fading channels for a  given outage probability requirement.
However, all these joint designs are offline and thus suffer from a substantial performance degradation when the adopted channel model diverges from the actual one encountered online.
As such, an online trajectory design has been proposed in \cite{JR:Yan_UAV_TCOM} to maximize the achievable throughput in the current time slot and the expected throughput in future time slots based on real-time and statistical knowledge of the channel gains, respectively.
Furthermore, a hybrid online and offline design was proposed in \cite{YouHybridOffOnLine}, where an offline trajectory design based on a probabilistic LoS channel model was followed by an online calibration exploiting the updated information from instantaneous channel measurements.
However, these works \cite{JR:Yan_UAV_TCOM,YouHybridOffOnLine} assumed perfect knowledge of channel state information (CSI) and did not consider channel estimation, which is generally needed in practice.

The second technical challenge of UAV-enabled communication is that it is highly susceptible to eavesdropping as the associated wiretap channels are also LoS-dominated in general\cite{QingqingPLSUAV}.
Therefore, the exchange of confidential information in UAV wireless networks has to be safeguarded.
As a remedy, the authors in \cite{QianPRLUAV} employed a buffer-aided UAV relay in the wiretap channel and proposed a secure resource allocation design to maximize the secrecy rate considering the information causality constraint at the UAV relay.
Also, in \cite{GuangchiUAV}, the authors proposed a joint trajectory and power control scheme to maximize the average secrecy rates by
exploiting the high mobility of the UAV to proactively establish favorable and degraded channels for the legitimate and eavesdropping links, respectively.
Furthermore, the authors in \cite{XiaoboUAV} considered a UAV base station serving multiple legitimate users with the aid of a UAV jammer and proposed a secure resource allocation design to maximize the minimum average secrecy rate among all users.
However, these secure resource allocation designs assumed the availability of the perfect CSI of the eavesdropping links \cite{QianPRLUAV,GuangchiUAV,XiaoboUAV}, which is an overly optimistic assumption in practice.
On the other hand, for the case of imperfect CSI, the authors of\cite{MiaoUAV} proposed a robust and secure resource allocation design to maximize the average worst-case secrecy rate of the system.
Also, based on the imperfect CSI of the eavesdroppers, the authors in \cite{CaiEEUAV} proposed to maximize the system energy efficiency given the limited energy storage capability of the onboard battery.
Yet, all these works \cite{QianPRLUAV,GuangchiUAV,XiaoboUAV,CaiEEUAV,MiaoUAV} assumed static eavesdroppers and their performances may degrade significantly when the eavesdropper is moving.
In practice, when the eavesdropper is maneuverable, such as an eavesdropping UAV (E-UAV), guaranteeing communication security becomes challenging due to time-varying CSI of the eavesdropping links.
However, conventional training-based channel estimation schemes are not applicable for wiretap channels as the eavesdropper does generally not cooperate with the legitimate transmitter and receiver.

Fortunately, electromagnetic (EM) waves can be exploited for both information acquisition and  delivery, which motivates the integration of sensing and communication\cite{FanISAC_tcom,AndrewCST}.
In fact, integrated sensing and communication (ISAC) has received considerable attention from both industry and academia for the shared use of the spectrum and the hardware platform as well as a joint signal processing framework.
Thus, in this paper, we propose to integrate sensing into UAV networks for the estimation of the state of an E-UAV, including its location and velocity, such that the wiretap channel can be inferred to facilitate secure wireless communications.
The general concept of sensing-aided physical layer security (PLS) has been recently discussed in \cite{wei2021towards}, but a corresponding practical design has not been proposed.
Also, the authors in \cite{su2021secure} studied PLS in dual-function radar-communication (DFRC) systems, where the multi-user interference was designed to be constructive at the legitimate users, while disrupting the eavesdropper.
Besides, assuming a static target, an integrated sensing, communication, and jamming scheme was proposed for DFRC systems for perfect and imperfect CSI of the wiretap channel in \cite{NanchiTWC}.
However, this scheme does not include the sensing of the wiretap channel and it is not applicable for mobile eavesdroppers.

In this paper, we propose an integrated sensing, navigation, and communication (ISNC) framework to guarantee PLS in UAV-enabled wireless networks in the presence of a mobile eavesdropper.
In particular, an information UAV (I-UAV) serving as an aerial base station transmits confidential messages to multiple legitimate ground users (GUs) and jams an E-UAV with artificial noise (AN) to facilitate secure communications.
At the same time, the I-UAV tracks the location and velocity of the E-UAV by exploiting the reflected AN employing an extended Kalman filter (EKF).
Based on the sensing information obtained in the previous time slot, the I-UAV designs its navigation strategy, predicts the wiretap channel parameters between I-UAV and E-UAV for the next time slot, and accordingly designs the communication resource allocation policy to optimize the secure communication and tracking performance in the next time slot.
A distance-based online navigation design is proposed to determine the flight direction and the velocity of the I-UAV for the next time slot.
Besides, we propose to employ a fully-connected neural network to model the wiretap channel prediction uncertainty to facilitate the robust resource allocation design.
Specifically, the proposed design optimizes the trade-off between sensing and communication in the next time slot, while taking into account the wiretap channel prediction error and the quality-of-service (QoS) requirements regarding the information leakage to the E-UAV and the achievable data rate of the legitimate GUs.
We develop a low-complexity suboptimal solution for online robust resource allocation design, which is based on a channel correlation-based user scheduling strategy and a semidefinite relaxation (SDR)-based precoding design.
Our simulation results demonstrate the benefits of integrating sensing and navigation for safeguarding UAV wireless networks.

The following notations are used in this paper. Boldface capital and lower case letters are reserved for matrices and vectors, respectively. $\mathbb{C}^{M\times N}$ denotes the set of all $M\times N$ matrices with complex entries; $\mathbb{R}^{M\times N}$ denotes the set of all $M\times N$ matrices with real entries; ${\left( \cdot \right)^{\mathrm{T}}}$ and ${\left( \cdot \right)^{\mathrm{H}}}$ denote the transpose and the Hermitian transpose of a vector or a matrix, respectively;
$\abs{\cdot}$ denotes the absolute value of a complex scalar;
and $\left\|\cdot\right\|$ denotes the Euclidean norm of a vector.
$\Tr\left( \cdot \right)$ denotes the trace of a matrix; $\diag\{\mathbf{x}\}$ denotes a diagonal matrix whose main diagonal elements are given by its input vector $\mathbf{x}$; $\mathbf{I}_{M}$ denotes the $M \times M$ identity matrix;
$\left\{\mathbf{X}\right\}_{ij}$ returns the entry in the $i$-th row and $j$-th column of matrix $\mathbf{X}$.
$\otimes$ and $\frac{\partial f(x)}{\partial x}$ denote the Kronecker product and the differential operator, respectively.
The real-valued Gaussian distribution with mean $\boldsymbol{\mu}$ and covariance matrix $\boldsymbol{\Sigma}$ is denoted by ${\cal N}(\boldsymbol{\mu},\boldsymbol{\Sigma})$, and the circularly symmetric complex Gaussian distribution with mean $\boldsymbol{\mu}$ and covariance matrix $\boldsymbol{\Sigma}$ is denoted by ${\cal CN}(\boldsymbol{\mu},\boldsymbol{\Sigma})$.

\vspace{-4mm}
\section{System Model and Proposed ISNC Framework}
\begin{figure}[t]
	\vspace{-5mm}
	\center{
		\includegraphics[width=3in]{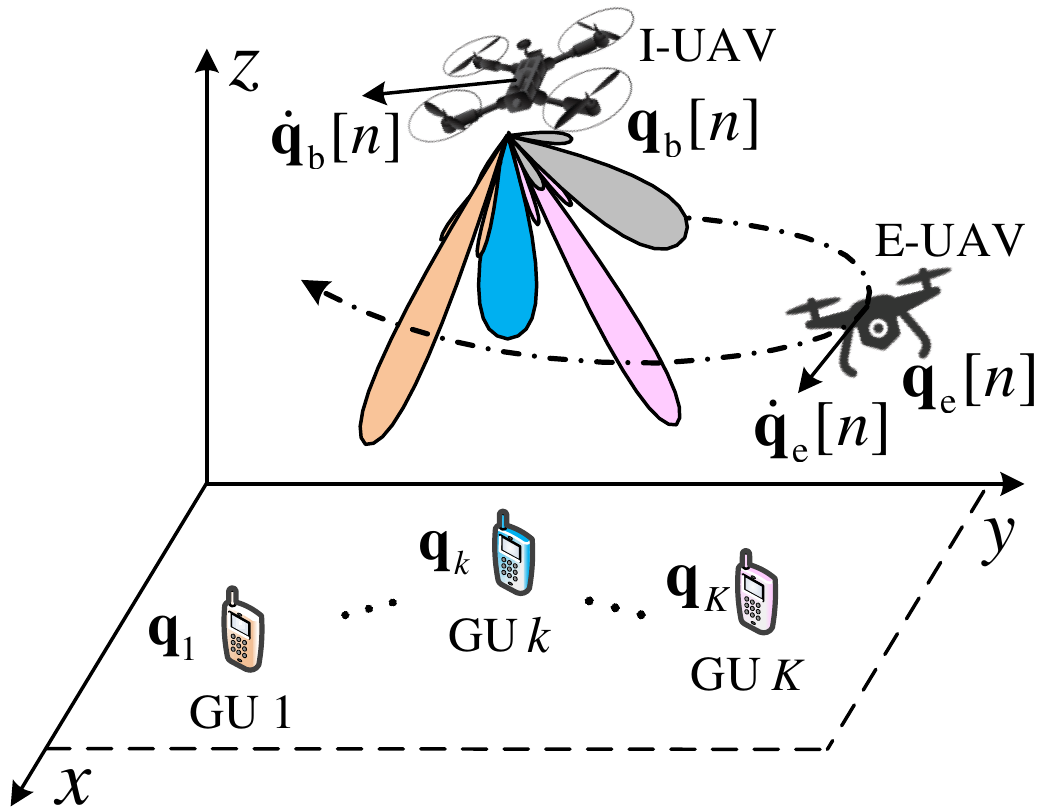}}\vspace{-7mm}
	\caption{A downlink I-UAV serving $K$ GUs in the presence of an E-UAV.}
	\label{fig:SystemModel}
	\vspace{-10mm}
\end{figure}

\vspace{-2mm}
\subsection{System Model}
\begin{figure}[t]
	\vspace{-5mm}
	\center{\includegraphics[width=4in]{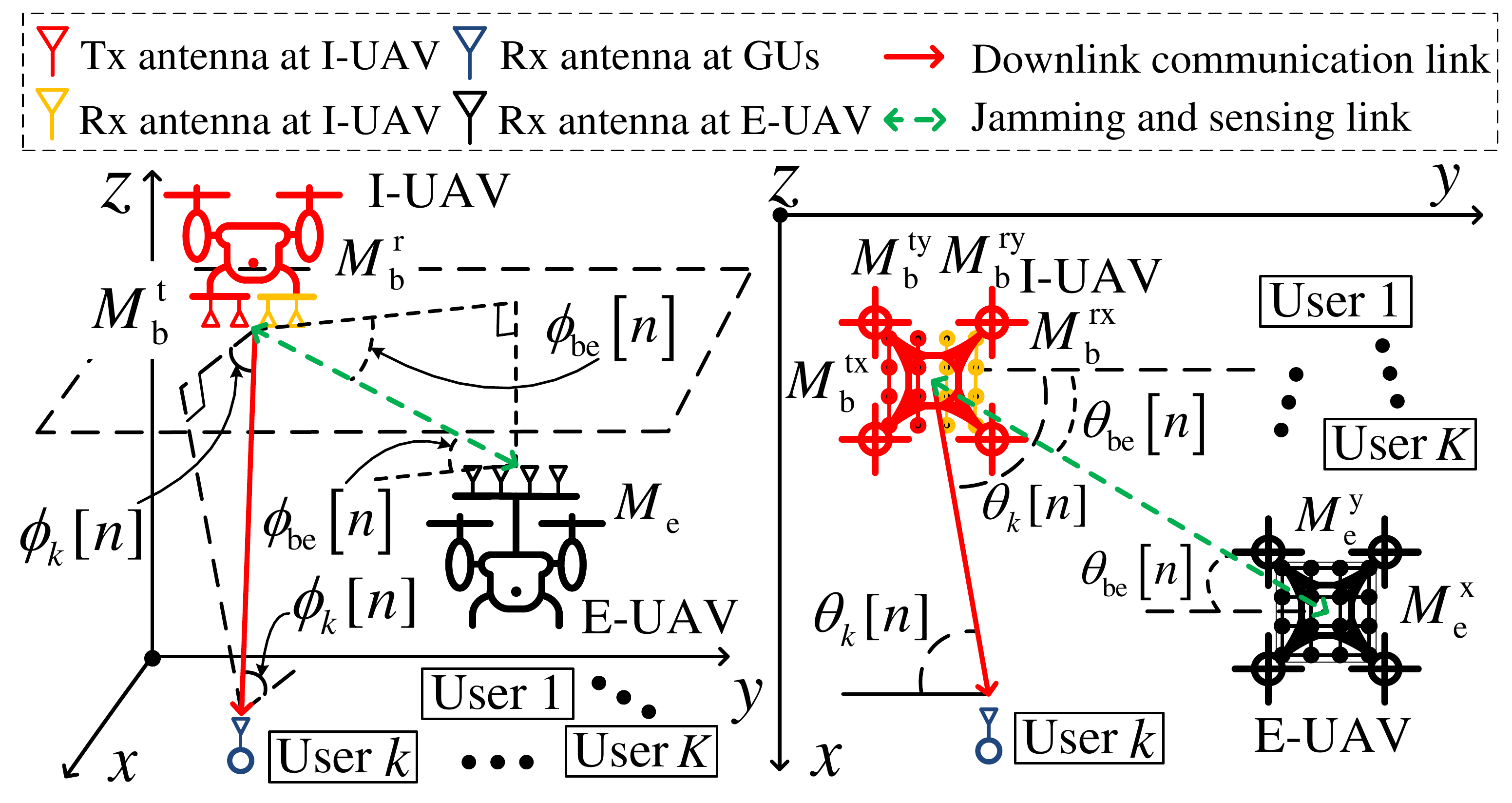}}\vspace{-7mm}
	\caption{The LoS channel model for the considered system.}
	\label{fig:AOD}
	\vspace{-7mm}
\end{figure}

We consider a downlink UAV communication system, where a mobile I-UAV serves as a base station opportunistically broadcasting $K$ independent confidential data streams to $K$ legitimate single-antenna GUs in the presence of a flying E-UAV, cf. Fig. \ref{fig:SystemModel}.
%
%
The I-UAV is equipped with two rectangular uniform planar arrays (UPAs) for three-dimensional (3D) transmit (Tx) and receive (Rx) beamforming, respectively, comprising
$M^{\mathrm{t}}_{\rm{b}}$ ($M^{\mathrm{tx}}_{\rm{b}}$ rows and $M^{\mathrm{ty}}_{\rm{b}}$ columns) and $M^{\mathrm{r}}_{\rm{b}}$ ($M^{\mathrm{rx}}_{\rm{b}}$ rows and $M^{\mathrm{ry}}_{\rm{b}}$ columns) antennas, respectively, where $M^{\mathrm{t}}_{\rm{b}} \ge K+1$, as illustrated in Fig. \ref{fig:AOD}.
The E-UAV is equipped with a Rx UPA, comprising $M_{\rm{e}}$ ($M^{\rm{x}}_{\rm{e}}$ rows and $M^{\rm{y}}_{\rm{e}}$ columns) antennas.
The total service time $T$ is divided into $N$ equal-length time slots with a slot duration of $\delta$, i.e., $T = N\delta$.
The E-UAV flies along a given trajectory, $\mathbf{q}_{\mathrm{e}}[n] = \left[x_{\mathrm{e}}[n],y_{\mathrm{e}}[n],z_{\mathrm{e}}[n]\right]^{\mathrm{T}}$, which is unknown to the I-UAV and is designed for intercepting the legitimate information transmission with velocity $\dot{\mathbf{q}}_{\mathrm{e}}[n] = [\dot{x}_{\mathrm{e}}[n],\dot{y}_{\mathrm{e}}[n],\dot{z}_{\mathrm{e}}[n]]^{\mathrm{T}}$, $\forall n \in \left\{1,\ldots,N\right\}$, where $\left[x_{\mathrm{e}}[n],y_{\mathrm{e}}[n],z_{\mathrm{e}}[n]\right]$ and $[\dot{x}_{\mathrm{e}}[n],\dot{y}_{\mathrm{e}}[n],\dot{z}_{\mathrm{e}}[n]]$ denote the Cartesian coordinates of the position and the velocity of the E-UAV in time slot $n$, respectively.
We assume that the E-UAV employs a fixed non-adaptive trajectory to cruise above the GUs, see Section \ref{Sec:SimlationResults}\footnote{In this paper, we consider a fixed non-adaptive E-UAV trajectory. Note that the E-UAV may also navigate to improve its interception capability based on sensing information it acquires online. Properly modeling this scenario requires a dynamic game-theoretic framework \cite{EldosoukyUAV}, which is an interesting extension of this work.}.
In time slot $n$, the I-UAV obtains its own position $\mathbf{q}_{\mathrm{b}}\left[n\right] = \left[x_{\mathrm{b}}\left[n\right],y_{\mathrm{b}}\left[n\right],z_{\mathrm{b}}\left[n\right]\right]^{\mathrm{T}}$ and velocity $\dot{\mathbf{q}}_{\mathrm{b}}[n] = [\dot{x}_{\mathrm{b}}[n],\dot{y}_{\mathrm{b}}[n],\dot{z}_{\mathrm{b}}[n]]^{\mathrm{T}}$ 
from its internal sensors and/or inertial navigation system (INS).
To facilitate the presentation, the states of the I-UAV and E-UAV are defined as $\boldsymbol{\alpha}_{\mathrm{b}} [n]  =  [\mathbf{q}^{\mathrm{T}}_{\mathrm{b}}[n], \dot{\mathbf{q}}^{\mathrm{T}}_{\mathrm{b}}[n]]^{\mathrm{T}}$ and $\boldsymbol{\alpha}_{\mathrm{e}} [n]  =  [\mathbf{q}^{\mathrm{T}}_{\mathrm{e}}[n], \dot{\mathbf{q}}^{\mathrm{T}}_{\mathrm{e}}[n]]^{\mathrm{T}}$, respectively.
All GUs are assumed to be fixed and their locations are given by $\mathbf{q}_k = \left[x_k,y_k,0\right]^{\mathrm{T}}$, $\forall k \in \mathcal{K} = \{1,\ldots,K\}$.
For the downlink communication design, the locations of the I-UAV, $\mathbf{q}_{\mathrm{b}}$, and the GUs, $\mathbf{q}_k$, $\forall k$, are assumed to be perfectly known and are shared between the nodes via separate feedback or feedforward links, as is commonly assumed in the literature \cite{QingqingMultiUAV,ZengAccessSky,Changsheng3DRician}.
The state of the E-UAV, $\boldsymbol{\alpha}_{\mathrm{e}} [n]$, is unknown to the I-UAV and will be sensed by the I-UAV online with our proposed scheme.
In contrast, the locations of both the I-UAV and GUs are assumed to be known at the E-UAV, which is the worst case scenario in terms of secure communication provisioning.

\begin{figure}[t]
	\center{\includegraphics[width=4.5in]{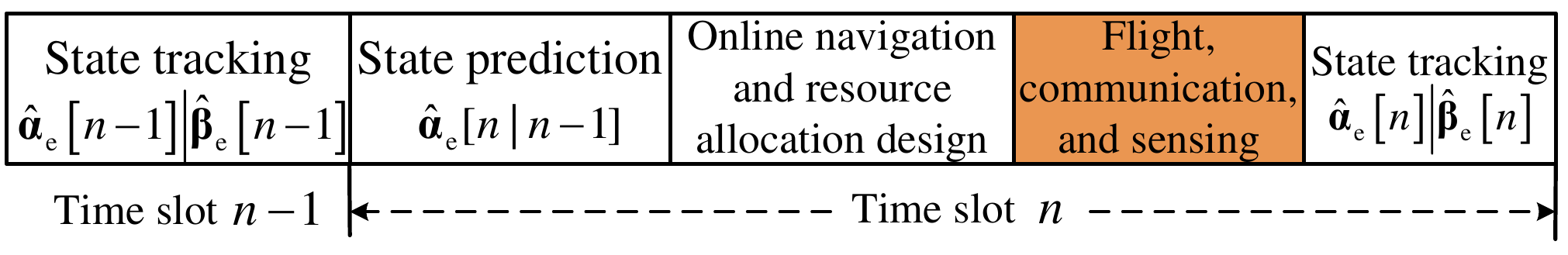}}\vspace{-7mm}
	\caption{The proposed ISNC protocol for UAV wireless networks.}
	\label{fig:SlotStructure}
	\vspace{-10mm}
\end{figure}

The proposed ISNC protocol is illustrated in Fig. \ref{fig:SlotStructure}, where $\hat{\boldsymbol{\beta}}_{\mathrm{e}} [n]$ represents the collection of all observable parameters of the E-UAV at the I-UAV in time slot $n$.
At the beginning of time slot $n$, the I-UAV first predicts the state of the E-UAV based on the state estimates in time slot $n-1$, i.e., $\hat{\boldsymbol{\alpha}}_{\mathrm{e}}[n|n-1]$, assuming a first-order Markov process for modeling the E-UAV movement.
Based on the predicted state, the I-UAV designs the navigation and resource allocation policy for time slot $n$.
Accordingly, the I-UAV flies to its designed location and broadcasts the information for the GUs and the AN adopting the resource allocation strategy designed for time slot $n$.
Exploiting the received echoes from the E-UAV, the I-UAV takes a new measurement $\hat{\boldsymbol{\beta}}_{\mathrm{e}} [n]$ and estimates the new state of the E-UAV $\hat{\boldsymbol{\alpha}}_{\mathrm{e}} [n]$, which is adopted as the input of the predictor for time slot $n+1$.
Note that in the considered system only causal information is available.
Thus, the adaptive navigation and resource allocation design in time slot $n$ has to be based on the measurements and state estimates from time slot $n-1$.
Hence, it is imperative to adopt an online design framework, which is different from the offline design-based UAV communication frameworks in the literature, e.g., \cite{ZhiqiangUAV,QianPRLUAV,GuangchiUAV,XiaoboUAV,CaiEEUAV,MiaoUAV}.
More details of the communication and tracking model adopted in the proposed ISNC framework are presented in the subsequent sections.

\vspace{-4mm}
\subsection{Communication Model}
In time slot $n$, the downlink transmit signal of the I-UAV is given by
\vspace{-1mm}
\begin{equation}\label{Eqn:TxSignal}
\mathbf{x}(n,t) = \sum\nolimits_{k = 1}^{K}  \mathbf{w}_k [n] u_k [n] s_k (n,t) + \mathbf{w}_{\mathrm{e}} [n]   a_{\mathrm{e}} (n,t),\vspace{-1mm}
\end{equation}
where $t \in \left(0,\delta\right)$ is a time instant within time slot $n$, $s_k (n,t) \sim \mathcal{CN} \left(0,1\right)$ denotes the transmitted information signal for GU $k$, and $a_{\mathrm{e}} (n,t) \sim \mathcal{CN} \left(0,1\right)$ is the AN for jamming.
The binary variable $u_{k}[n] = 1$ indicates that GU $k$ is selected for communication in time slot $n$, otherwise, $u_{k}[n] = 0$.
Vectors $\mathbf{w}_k [n] \in \mathbb{C}^{M^{\mathrm{t}}_{\mathrm{b}} \times 1}$ and $\mathbf{w}_{\mathrm{e}} [n] \in \mathbb{C}^{M^{\mathrm{t}}_{\mathrm{b}} \times 1}$ are the precoding vectors for GU $k$ and AN in time slot $n$, respectively.
The received signal at GU $k$ in time slot $n$ and time instant $t$ is given by
\vspace{-2mm}
\begin{equation}\label{Eqn:ReceivedSignalModelI}
y_k(n,t) = e^{j2\pi\nu_{k}[n]t}\mathbf{h}^{\mathrm{H}}_{k}[n] \mathbf{x}(n,t-\tau_{k}[n]) + v_k(n,t), \vspace{-2mm}
\end{equation}
where ${v}_k(n,t) \sim \mathcal{CN}\left(0,\sigma_k^2\right)$ denotes the additive white Gaussian noise (AWGN) at GU $k$ with power $\sigma_k^2$.
Variables $\tau_{k}[n]$ and  $\nu_{k}[n]$ denote the time delay and Doppler shift for the link from the I-UAV to GU $k$ in time slot $n$, respectively.
As this is the first work on ISNC for UAV wireless networks, to gain insights for system design, we assume pure LoS channels.
Thus, the channel vector between the I-UAV and GU $k$ in time slot $n$ is given by 
\vspace{-1mm}
\begin{equation}\label{Eqn:ReceivedSignalModelI_LoS}
\mathbf{h}_{k}[n] ={\frac{\beta_0}{d_k[n]}} \mathbf{a}_{M_{\mathrm{b}}^{\mathrm{tx}},M_{\mathrm{b}}^{\mathrm{ty}}}\left(\theta_{k}[n],\phi_{k}[n]\right),\vspace{-1mm}
\end{equation}
where $\beta^2_0$ represents the channel power gain at a reference distance and $d_k[n] = \|\mathbf{q}_{\mathrm{b}}[n] - \mathbf{q}_{k}[n]\|$ denotes the distance between the I-UAV and GU $k$.
	%
%
Angles $\theta_{k}[n]\in \left[-\pi/2,\pi/2\right]$ and $\phi_{k}[n] \in \left[-\pi/2,0\right)$ denote the azimuth angle of departure (AOD)\footnote{In this paper, we assume that array signal processing approaches \cite{van2004optimum,NairOCEANS} can address the 
	left-right ambiguity issue of UPAs and thus the azimuth angle can be assumed to be within a half angular space, i.e., $\theta_{k}[n]\in \left[-\pi/2,\pi/2\right]$.}  and the elevation AOD from the I-UAV to GU $k$ in time slot $n$, respectively, cf. Fig. \ref{fig:AOD}.
$\mathbf{a}_{M^{\mathrm{x}},M^{\mathrm{y}}}\left(\theta, \phi\right) \in \mathbb{C}^{M^{\mathrm{x}}M^{\mathrm{y}} \times 1}$ is the steering vector of a UPA with $M^{\mathrm{x}}$ and $M^{\mathrm{y}}$ antennas per row and column, respectively, and it is given by\cite{van2004optimum}
\vspace{-1mm}
\begin{align}\label{Eqn:UPASteeringVector}
\mathbf{a}_{M^{\mathrm{x}},M^{\mathrm{y}}}\left(\theta, \phi\right) &= \left( 1, e^{-  \frac{j2 \pi d\cos \phi \sin \theta}{\lambda_{\mathrm c}} } , \ldots, e^{-  \frac{j2(M^{\mathrm{x}}-1) \pi d\cos \phi \sin \theta}{\lambda_{\mathrm c}}     } \right)^{\mathrm T} \notag\\[-1mm]
&\otimes \left( 1, e^{-  \frac{j2 \pi d\cos \phi \cos \theta}{\lambda_{\mathrm c}}  } , \ldots, e^{-  \frac{j2 ({M^{\mathrm{y}}}-1) \pi d\cos \phi\cos \theta}{\lambda_{\mathrm c}}  } \right)^{\mathrm T},
\end{align}
\vspace{-8mm}\par\noindent
where $d$ denotes the spacing of neighboring antennas in a row or column of the I-UAV UPA, and ${\lambda_{\mathrm c}} $ is the wavelength of the transmit signal.
Assuming perfect time and frequency synchronization to compensate for the time delay and Doppler shift at all GUs \cite{Nasir2016}, the resulting received signal at GU $k$ in time slot $n$ and time instant $t$ is given by
\vspace{-1mm}
\begin{align}\label{Eqn:ReceivedSignalModelII}
\widetilde{y}_k(n,t) &= \underbrace {\mathbf{h}^{\mathrm{H}}_{k}[n] \mathbf{w}_k [n] u_k [n] s_k (n,t)}_{\text{Desired signal}}  + \underbrace{\sum\nolimits_{k' \neq k}  \mathbf{h}^{\mathrm{H}}_{k}[n]\mathbf{w}_{k'} [n] u_{k'} [n] s_{k'} (n,t)}_{\text{Inter-user interference}} \notag\\[-2mm]
&+ \underbrace{ \mathbf{h}^{\mathrm{H}}_{k}[n]\mathbf{w}_{\mathrm{e}} [n] a_{\mathrm{e}} (n,t)}_{\text{AN}} + \widetilde{v}_k(n,t),
\end{align}
\vspace{-5mm}\par\noindent
where $\widetilde{v}_k(n,t) \in \mathcal{CN}\left(0,\sigma_k^2\right)$ denotes the AWGN after time and frequency synchronization.
Note that the noise power is not affected by perfect time and frequency synchronization\cite{Nasir2016}.
In addition to AWGN, the received signal of GU $k$ is impaired by inter-user interference (IUI) and AN, i.e., the second and third terms in \eqref{Eqn:ReceivedSignalModelII}, respectively.
As a result, the achievable data rate of GU $k$ in time slot $n$ is given by
\vspace{-1mm}
\begin{equation}\label{Eqn:DownlinkRate}
{R_k}[n] = {\log _2}\Bigg({1 + \frac{{{{u_{k}[n]\big| {{\bf{h}}_{k}^{\rm{H}}[n]{{\bf{w}}_k}[n]} \big|}^2}}}{{\sum\limits_{k' \neq k} {{{u_{k'}[n]\big| {{\bf{h}}_{k}^{\rm{H}}[n]{{\bf{w}}_{k'}}[n]} \big|}^2}}  + {{\big| {{\bf{h}}_{k}^{\rm{H}}[n]{{\bf{w}}_{\rm{e}}}[n]} \big|}^2} + {{\sigma _k^2}} }}} \Bigg).\vspace{-1mm}
\end{equation}

The received signal at the E-UAV in time slot $n$ at time instant $t$ is given by
\vspace{-1mm}
\begin{equation}\label{Eqn:ReceivedSignalEUAV_I}
\mathbf{y}_{\mathrm{e}}(n,t) = e^{j2\pi\nu_{\mathrm{be}}[n]t} \mathbf{H}_{\mathrm{e}}[n] \mathbf{x}(n,t-\tau_{\mathrm{be}}[n]) + \mathbf{v}_{\mathrm{e}}(n,t),\vspace{-1mm}
\end{equation}
where $\mathbf{v}_{\mathrm{e}}(n,t) \in \mathcal{CN}\left(\mathbf{0},\sigma_{\mathrm{e}}^2\mathbf{I}_{M_{\mathrm{e}}}\right)$ denotes the background noise at the E-UAV with power $\sigma_{\mathrm{e}}^2$, and $\tau_{\mathrm{be}}[n]$ and $\nu_{\mathrm{be}}[n]$ denote respectively the time delay and Doppler shift for the link between I-UAV and E-UAV.
Matrix $\mathbf{H}_{\mathrm{e}}[n] \in \mathbb{C}^{M_{\mathrm{e}} \times M^{\mathrm{t}}_{\mathrm{b}}}$ represents the channel matrix between I-UAV and E-UAV and it is given by
\vspace{-2mm}
\begin{equation}\label{Eqn:ChannelMatrix}
\mathbf{H}_{\mathrm{e}}[n] = \frac{\beta_0}{d_{\mathrm{e}}[n]} \mathbf{a}_{M_{\mathrm{e}}^{\mathrm{x}},M_{\mathrm{e}}^{\mathrm{y}}}\left(\theta_{\mathrm{eb}}[n], \phi_{\mathrm{eb}}[n]\right)\mathbf{a}^{\mathrm{H}}_{{M^{\mathrm{tx}}_{\mathrm{b}}},{M^{\mathrm{ty}}_{\mathrm{b}}}}\left(\theta_{\mathrm{be}}[n], \phi_{\mathrm{be}}[n]\right),\vspace{-1mm}
\end{equation}
where $d_{\mathrm{e}}[n] = \|\mathbf{q}_{\mathrm{b}}[n] - \mathbf{q}_{{\mathrm{e}}}[n]\|$ is the distance between I-UAV and E-UAV.
Angles $\theta_{\mathrm{be}}[n]\in \left[-\pi/2,\pi/2\right]$ and  $\phi_{\mathrm{be}}[n]\in \left[-\pi/2,\pi/2\right]$ denote the azimuth and elevation AODs from the I-UAV to the E-UAV in time slot $n$, respectively, cf. Fig. \ref{fig:AOD}.
Also, as shown in Fig. \ref{fig:AOD}, angles $\theta_{\mathrm{eb}}[n]\in \left[-\pi/2,\pi/2\right]$ and  $\phi_{\mathrm{eb}}[n]\in \left[-\pi/2,\pi/2\right]$ denote the azimuth angle of arrival (AOA) and the elevation AOA from the I-UAV to the E-UAV in time slot $n$, respectively.
Note that apart from the AODs and AOAs, the channel between I-UAV and E-UAV depends on their respective 3D orientations, including their yaw, pitch, and roll angles\cite{DongfangXu3DUAV}.
Here, we assume that the 3D orientations of I-UAV and E-UAV can be acquired by their internal gyroscope sensors and be compensated via adjusting the designed transmit and receive beamforming vectors according to the yaw, pitch, and roll angles.
Therefore, the 3D orientations do not affect our proposed precoding design and thus are not explicitly included in $\mathbf{H}_{\mathrm{e}}[n]$ in \eqref{Eqn:ChannelMatrix}.

As the precoding policy of the I-UAV is unknown to the E-UAV, we assume that the E-UAV performs maximum ratio combining (MRC) to maximize its received signal power\footnote{We note that maximizing the received signal power of GU $k$ while minimizing the AN power is the optimal strategy for intercepting the information of GU $k$. However, this requires the E-UAV to be aware of the precoding vectors employed in each time slot at the I-UAV for GU $k$ and the AN, which might be impossible in practice. Therefore, we assume the E-UAV adopts MRC to facilitate resource allocation design. The extension to other combining strategies is an interesting topic for future work.},  i.e., the receive beamforming vector at the E-UAV is given by $\mathbf{u}_{\mathrm{e}}[n] = \sqrt{\frac{1}{M_{\mathrm{e}}}} \mathbf{a}_{M_{\mathrm{e}}^{\mathrm{x}},M_{\mathrm{e}}^{\mathrm{y}}}\left(\theta_{\mathrm{eb}}[n], \phi_{\mathrm{eb}}[n]\right)$, where $\theta_{\mathrm{eb}}[n]$, $\phi_{\mathrm{eb}}[n]$, and $d_{\mathrm{e}}[n]$ can be inferred from the locations of both I-UAV and E-UAV.
Furthermore, assuming perfect time and frequency synchronization at the E-UAV, the resulting received signal at the E-UAV in time slot $n$ and time instant $t$ is given by
\vspace{-1mm}
\begin{equation}\label{Eqn:ReceivedSignalEUAV_II}
\widetilde{y}_{\mathrm{e}}(n,t) = \underbrace{\sum\nolimits_{k = 1}^{K} \sqrt{M_{\mathrm{e}}} \mathbf{h}^{\mathrm{H}}_{\mathrm{be}}[n] \mathbf{w}_k [n] u_{k}[n] s_k (n,t)}_{\text{Downlink information}} +\underbrace{\sqrt{M_{\mathrm{e}}}{\mathbf{h}^{\mathrm{H}}_{\mathrm{be}}[n]\mathbf{w}_{\mathrm{e}} [n] a_{\mathrm{e}} (n,t)}}_{\text{AN}} + \widetilde{v}_{\mathrm{e}}(n,t),
\end{equation}
where $\widetilde{v}_{\mathrm{e}}(n,t) \in \mathcal{CN}\left(0,\sigma_{\mathrm{e}}^2\right)$ denotes the post-processing noise and the effective wiretap channel between I-UAV and E-UAV, $\mathbf{h}_{\mathrm{be}}[n]$, is given by 
\begin{equation}\label{Eqn:EffectiveChannel}
	\mathbf{h}_{\mathrm{be}}[n] = \frac{\beta_0}{d_{\mathrm{e}}[n]}\mathbf{a}_{{M^{\mathrm{tx}}_{\mathrm{b}}},{M^{\mathrm{ty}}_{\mathrm{b}}}}\left(\theta_{\mathrm{be}}[n], \phi_{\mathrm{be}}[n]\right).
\end{equation}
For intercepting the information of GU $k$, as a worst case, we assume that the E-UAV can mitigate the IUI caused by other GUs' signal, such that the AN is the only impairment, which is a common worst case assumption in the literature\cite{Mukherjee2014}.
Thus, the leakage information rate associated with GU $k$ in time slot $n$ is given by
\vspace{-2mm}
\begin{equation}\label{Eqn:EavesdroppingUAVRate}
{R^{k}_{\rm{e}}}[n] = {\log _2}\Bigg( {1 + \frac{{{{u_{k}[n]M_{\rm{e}}\big| {{\bf{h}}_{{\rm{be}}}^{\rm{H}}[n]{{\bf{w}}_k}[n]} \big|}^2}}}{{{{M_{\rm{e}}\big| {{\bf{h}}_{{\rm{be}}}^{\rm{H}}[n]{{\bf{w}}_{\rm{e}}}[n]} \big|}^2} + {\sigma _{\rm{e}}^2}}}} \Bigg).\vspace{-4mm}
\end{equation}


\vspace{-4mm}
\subsection{Tracking Model}
\label{Subsec:Tracking Model}

\subsubsection{Echo Signal Model}
In practice, the signal transmitted by the I-UAV is partially received by the receive antennas of the E-UAV and is partially reflected by the latter's body.
The I-UAV is assumed to operate in the full-duplex mode, which allows it to transmit and receive signals simultaneously in the same frequency band, while the received signal may suffer residual self-interference\cite{SunYanNOMA}.
The echo signal received at the I-UAV in time slot $n$ is given by
\vspace{-1mm}
\begin{align}\label{Eqn:EchoesUAV}
\mathbf{r}_{\mathrm{b}}(n,t)&= \underbrace{ e^{j2\pi\nu_{\mathrm{e}}[n]t}\mathbf{H}^{\mathrm{r}}_{\mathrm{be}}[n] \sum\nolimits_{k = 1}^{K}  \mathbf{w}_k [n] u_k[n]s_k (n,t-\tau_{\mathrm{e}}[n])}_{\text{Echo of downlink information signal}}  \notag\\[-2mm]
&+ \underbrace{ e^{j2\pi\nu_{\mathrm{e}}[n]t}\mathbf{H}^{\mathrm{r}}_{\mathrm{be}}[n] \mathbf{w}_{\mathrm{e}} [n] a_{\mathrm{e}} (n,t-\tau_{\mathrm{e}}[n])}_{\text{Echo of AN}} + \mathbf{v}_{\mathrm{b}}(n,t),
\end{align}
\vspace{-5mm}\par\noindent
where the round-trip channel matrix $\mathbf{H}^{\mathrm{r}}_{\mathrm{be}}[n] \in \mathbb{C}^{M^{\mathrm{r}}_{\mathrm{b}} \times M^{\mathrm{t}}_{\mathrm{b}}}$ is given by
\vspace{-1mm}
\begin{equation}\label{Eqn:RoundTripChannel}
\mathbf{H}^{\mathrm{r}}_{\mathrm{be}}[n] =\frac{\epsilon_{\mathrm{e}}[n]\beta_0}{2d_{\mathrm{e}}[n]}  \mathbf{a}_{M_{\mathrm{b}}^{\mathrm{rx}},M_{\mathrm{b}}^{\mathrm{ry}}}\left(\theta_{\mathrm{be}}[n], \phi_{\mathrm{be}}[n]\right)\mathbf{a}^{\mathrm{H}}_{{M^{\mathrm{tx}}_{\mathrm{b}}},{M^{\mathrm{ty}}_{\mathrm{b}}}}\left(\theta_{\mathrm{be}}[n], \phi_{\mathrm{be}}[n]\right).\vspace{-1mm}
\end{equation}
Here, variables $\tau_{\mathrm{e}}[n]$ and $\nu_{\mathrm{e}}[n]$ denote the round-trip time delay and Doppler shift, respectively, $\epsilon_{\mathrm{e}}[n] = \sqrt{\frac{\vartheta_{\mathrm{e}}}{4\pi d^2_{\mathrm{e}}[n]}}$ denotes the reflection coefficient of the E-UAV in time slot $n$, and $\vartheta_{\mathrm{e}}$ is the radar
cross-section of the E-UAV \cite{skolnik1962introduction}.
Vector $\mathbf{v}_{\mathrm{b}}(n,t) \in \mathcal{CN} \left(\mathbf{0},\sigma_{\mathrm{b}}^2\mathbf{I}_{M^{\mathrm{r}}_{\mathrm{b}}}\right)$ captures both the background noise and the residual self-interference \cite{SunYanNOMA} at the I-UAV, where power $\sigma_{\mathrm{b}}^2$ is assumed to be constant for a given transmit power.
Note that clutter, such as signals reflected by GUs and other scatters in the environment, is omitted here as it can be effectively suppressed by existing clutter suppression techniques \cite{skolnik1962introduction} owing to its distinctive reflection angles and Doppler frequencies compared to the echoes received from the E-UAV \cite{skolnik1962introduction}.
Besides, we assume that the AOAs from the E-UAV to I-UAV are identical to the corresponding AODs from the I-UAV to E-UAV in \eqref{Eqn:RoundTripChannel}, which is reasonable for point target models \cite{FanLiuISAC} and reciprocal propagation channels.
With this assumption, the round-trip channel matrix depends on the azimuth and elevation AODs from the I-UAV to E-UAV.
In the sequel, we use $\theta_{\mathrm{e}}[n] = \theta_{\mathrm{be}}[n]$ and $\phi_{\mathrm{e}}[n] = \phi_{\mathrm{be}}[n]$  for concise notation.
Due to the unknown location of the E-UAV, $\tau_{\mathrm{e}}[n]$, $\nu_{\mathrm{e}}[n]$, $\theta_{\mathrm{e}}[n]$, and $\phi_{\mathrm{e}}[n]$ are \textit{a priori} unknown but can be estimated by the I-UAV based on the received echo signals as explained in the following.

\subsubsection{E-UAV Parameter Estimation Model}
Based on the echo signal in \eqref{Eqn:EchoesUAV}, different methods can be employed to estimate $\tau_{\mathrm{e}}\hspace{-0.25mm}[\hspace{-0.25mm}n\hspace{-0.25mm}]$, $\nu_{\mathrm{e}}\hspace{-0.25mm}[\hspace{-0.25mm}n\hspace{-0.25mm}]$, $\phi_{\mathrm{e}}\hspace{-0.25mm}[\hspace{-0.25mm}n\hspace{-0.25mm}]$, and $\theta_{\mathrm{e}}\hspace{-0.25mm}[\hspace{-0.25mm}n\hspace{-0.25mm}]$ \cite{kay1993fundamentals}.
One possible approach to estimate these parameters is the matched-filter (MF) principle by exploiting the reflected AN \cite{kay1993fundamentals}:
\vspace{-1mm}
\begin{equation}\label{Eqn:MF_Estimation}
\left\{\hspace{-0.5mm} {{\hat \tau _{\rm{e}}}[n],{\hat\nu _{\rm{e}}}[n]}\hspace{-0.5mm},\hat \theta_{\mathrm{e}}[n],\hat\phi_{\mathrm{e}}[n] \right\}\hspace{-1mm}  = \arg \mathop {\max }\limits_{\tau ,\nu, \theta, \phi} \hspace{-0.5mm}\Big|\hspace{-0.5mm} \frac{1}{\delta}\hspace{-0.5mm}{\int_0^\delta\hspace{-2mm} \mathbf{a}^{\mathrm{H}}_{M_{\mathrm{b}}^{\mathrm{rx}},M_{\mathrm{b}}^{\mathrm{ry}}}\hspace{-0.5mm}\left(\hspace{-0.5mm}\theta,\hspace{-0.5mm}\phi\hspace{-0.5mm}\right)\hspace{-0.5mm} {{{\bf{r}}_{\rm{b}}}\hspace{-0.5mm}(\hspace{-0.5mm}n,\hspace{-0.5mm}t\hspace{-0.5mm})a_{\rm{e}}^ *\hspace{-0.5mm} (\hspace{-0.5mm}n,\hspace{-0.5mm}t \hspace{-0.5mm}- \hspace{-0.5mm}\tau \hspace{-0.5mm}){e^{ - j2\pi \nu t}}dt} } \Big|.\vspace{-1mm}
\end{equation}
Note that we exploit the AN rather than the downlink information signal for sensing as adopting the latter would require the I-UAV to transmit the information-bearing signals towards the E-UAV, which increases the risk of potential information leakage.
Instead, exploiting the AN for both sensing and jamming is a win-win strategy for secrecy applications.
Analyzing the estimation variances associated with \eqref{Eqn:MF_Estimation} is a challenging task.
According to \cite{FanLiuISAC,kay1993fundamentals}, since the AN and the information-bearing signals are uncorrelated, i.e., $\frac{1}{\delta}\hspace{-1mm}\int_0^\delta\hspace{-0.5mm}  s_k\hspace{-0.25mm}(\hspace{-0.25mm}n,\hspace{-0.25mm}t\hspace{-0.25mm}) a^*_{\mathrm{e}}\hspace{-0.25mm}(\hspace{-0.25mm}n,\hspace{-0.25mm}t\hspace{-0.25mm}) dt \hspace{-0.5mm} \approx \hspace{-0.5mm} 0$, $\forall k$, the estimation variances of $\tau_{\mathrm{e}}[n]$, $\nu_{\mathrm{e}}[n]$, $\phi_{\mathrm{e}}[n]$, and $\theta_{\mathrm{e}}[n]$ are inversely proportional to the MF output signal-to-noise ratio (SNR) and can be modeled as $\sigma^2_{\tau _{\rm{e}}[n]} \hspace{-1mm}=\hspace{-1mm} c_{{\tau _{\rm{e}}}}/\mathrm{SNR}[n]$, $\sigma^2_{\nu _{\rm{e}}[n]} \hspace{-1mm}=\hspace{-1mm} c_{{\nu _{\rm{e}}}}/\mathrm{SNR}[n]$, $\sigma^2_{\theta_{\mathrm{e}}[n]} \hspace{-1mm}= \hspace{-1mm}c_{{\theta _{\rm{e}}}}/\mathrm{SNR}[n]$, and $\sigma^2_{ \phi_{\mathrm{e}}[n]} \hspace{-1mm}=\hspace{-1mm} c_{{\phi _{\rm{e}}}}/\mathrm{SNR}[n]$, respectively, where the MF output SNR in time slot $n$ is given by
\vspace{-1mm}
\begin{equation}\label{Eqn:MFsnr}
\mathrm{SNR}[n] \hspace{-1mm}=\hspace{-1mm} \frac{\vartheta_{\mathrm{e}}\beta^2_0G_{\mathrm{MF}}M^{\mathrm{r}}_{\mathrm{b}} \big|\mathbf{a}^{\mathrm{H}}_{{M^{\mathrm{tx}}_{\mathrm{b}}},{M^{\mathrm{ty}}_{\mathrm{b}}}}\left(\theta_{\mathrm{e}}[n], \phi_{\mathrm{e}}[n]\right) \mathbf{w}_{\mathrm{e}} [n]\big|^2}{16\pi \sigma_{\mathrm{b}}^2d^4_{\mathrm{e}}[n]}.\vspace{-1mm}
\end{equation}
The modeling parameters $c_{{\tau _{\rm{e}}}}, c_{\nu _{\rm{e}}}, c_{\theta_{\mathrm{be}}}, c_{\phi_{\mathrm{be}}}\hspace{-1mm}>\hspace{-1mm}0$ are assumed to be known and depend on the specific estimation method adopted \cite{FanLiuISAC}.
Furthermore, $G_{\mathrm{MF}}$ in \eqref{Eqn:MFsnr} is the MF gain, which is proportional to the number of symbols transmitted in one time slot.





\subsubsection{Measurement Model for the E-UAV}
The measurement model characterizes the relationship between the observable E-UAV parameters and the hidden E-UAV state, and is the key for inferring the state of the E-UAV.
%
%
In particular, considering the positions of I-UAV and E-UAV in Fig. \ref{fig:AOD}, the measurement models associated with $\tau_{\mathrm{e}}\hspace{-0.25mm}[\hspace{-0.25mm}n\hspace{-0.25mm}]$, $\nu_{\mathrm{e}}\hspace{-0.25mm}[\hspace{-0.25mm}n\hspace{-0.25mm}]$, $\phi_{\mathrm{e}}\hspace{-0.25mm}[\hspace{-0.25mm}n\hspace{-0.25mm}]$, and $\theta_{\mathrm{e}}\hspace{-0.25mm}[\hspace{-0.25mm}n\hspace{-0.25mm}]$ are given by
\begin{align}\label{Eqn:MeasurementModel_I}
{\tau} _{\rm{e}}[n] &\hspace{-1mm}=\hspace{-1mm} \frac{2\|\mathbf{q}_{\mathrm{e}} [n] - \mathbf{q}_{{\mathrm{b}}}[n]\|}{c} + v_{\tau_{{\mathrm{e}}}[n]},\notag\\[-1mm]
{{\nu} _{\rm{e}}}[n] &\hspace{-1mm}=\hspace{-1mm} \frac{2\left(\dot{\mathbf{q}}^{\mathrm{T}}_{\mathrm{e}} [n]-\dot{\mathbf{q}}^{\mathrm{T}}_{\mathrm{b}} [n]\right) \left(\mathbf{q}_{\mathrm{e}} [n] \hspace{-1mm}-\hspace{-1mm} \mathbf{q}_{{\mathrm{b}}}[n]\right) f_c}{c\|\mathbf{q}_{\mathrm{e}} [n] - \mathbf{q}_{{\mathrm{b}}}[n]\|} + v_{\nu _{\rm{e}}[n]},\notag\\[-1mm]
\sin {\theta}_{\mathrm{e}}[n] &\hspace{-1mm}=\hspace{-1mm} \frac{x_{\mathrm{e}}[n] \hspace{-1mm}-\hspace{-1mm} x_{\mathrm{b}}[n]}{\sqrt{\left|x_{\mathrm{e}}[n] \hspace{-1mm}-\hspace{-1mm} x_{\mathrm{b}}[n]\right|^2\hspace{-1mm}+\hspace{-1mm}\left|y_{\mathrm{e}}[n] \hspace{-1mm}-\hspace{-1mm} y_{\mathrm{b}}[n]\right|^2}} \hspace{-1mm}+\hspace{-1mm} v_{\sin \theta_{\mathrm{e}}[n]},\;\text{and}\notag\\[-1mm]
\sin {\phi}_{\mathrm{e}}[n] &\hspace{-1mm}=\hspace{-1mm} \frac{z_{\mathrm{e}}[n] - z_{\mathrm{b}}[n]}{\|\mathbf{q}_{\mathrm{e}} [n] - \mathbf{q}_{{\mathrm{b}}}[n]\|} + v_{\sin \phi_{\mathrm{e}}[n]},
\end{align}
\vspace{-7mm}\par\noindent
respectively, where $c$ is the speed of light and $f_c = c/\lambda_{\mathrm c}$ is the carrier frequency. 
Gaussian random variables $v_{\tau_{{\mathrm{e}}}[\hspace{-0.25mm}n\hspace{-0.25mm}]}$, $v_{\nu _{\rm{e}}[\hspace{-0.25mm}n\hspace{-0.25mm}]}$, $v_{\sin \theta_{\mathrm{e}}[\hspace{-0.25mm}n\hspace{-0.25mm}]}$, and $v_{\sin \phi_{\mathrm{e}}[\hspace{-0.25mm}n\hspace{-0.25mm}]}$ denote the corresponding measurement noises with zero mean and variances $\sigma^2_{\tau _{\rm{e}}[\hspace{-0.25mm}n\hspace{-0.25mm}]}$, $\sigma^2_{\nu _{\rm{e}}[\hspace{-0.25mm}n\hspace{-0.25mm}]}$, $\sigma^2_{\sin \theta_{\mathrm{e}}[\hspace{-0.25mm}n\hspace{-0.25mm}]}$, and $\sigma^2_{\sin \phi_{\mathrm{e}}[\hspace{-0.25mm}n\hspace{-0.25mm}]}$, respectively.
In the high SNR regime, i.e., $\sigma^2_{\theta_{\mathrm{e}}[\hspace{-0.25mm}n\hspace{-0.25mm}]},\sigma^2_{\phi_{\mathrm{e}}[\hspace{-0.25mm}n\hspace{-0.25mm}]}\hspace{-1mm}\to\hspace{-1mm} 0$, and using trigonometric identities, we have $\sigma_{\sin \theta_{\mathrm{e}}[\hspace{-0.25mm}n\hspace{-0.25mm}]} \hspace{-1mm}\approx\hspace{-1mm} \cos {\theta}_{\mathrm{e}}[\hspace{-0.25mm}n\hspace{-0.25mm}] \sigma_{\theta_{\mathrm{e}}[\hspace{-0.25mm}n\hspace{-0.25mm}]}$ and $\sigma_{\sin \phi_{\mathrm{e}}[\hspace{-0.25mm}n\hspace{-0.25mm}]} \hspace{-1mm}\approx\hspace{-1mm} \cos {\phi}_{\mathrm{e}}[\hspace{-0.25mm}n\hspace{-0.25mm}] \sigma_{\phi_{\mathrm{e}}[\hspace{-0.25mm}n\hspace{-0.25mm}]}$.

Now, we collect all the observable parameters in $\boldsymbol{\beta}_{\mathrm{e}} [\hspace{-0.25mm}n\hspace{-0.25mm}] \hspace{-1mm}= \hspace{-1mm}[{\tau} _{\rm{e}}[\hspace{-0.25mm}n\hspace{-0.25mm}],{{\nu} _{\rm{e}}}[\hspace{-0.25mm}n\hspace{-0.25mm}], \sin {\theta}_{\mathrm{e}}[\hspace{-0.25mm}n\hspace{-0.25mm}], \sin {\phi}_{\mathrm{e}}[\hspace{-0.25mm}n\hspace{-0.25mm}]]^\mathrm{T} \hspace{-1mm}\in\hspace{-1mm} \mathbb{R}^{4 \times 1}$. Then, the measurement model can be rewritten as
\vspace{-2mm}
\begin{equation}\label{Eqn:MeautrementModel_II}
\boldsymbol{\beta}_{\mathrm{e}} [n] =  \mathbf{g}_n\left(\boldsymbol{\alpha}_{\mathrm{e}} [n]\right) + \mathbf{v}_{\boldsymbol{\beta}_{\mathrm{e}}[n]},\vspace{-2mm}
\end{equation}
where $\mathbf{v}_{\boldsymbol{\beta}_{\mathrm{e}}[n]} = \left[v_{\tau_{{\mathrm{e}}}[n]},v_{\nu _{\rm{e}}[n]},v_{\sin \theta_{\mathrm{e}}[n]},v_{\sin \phi_{\mathrm{e}}[n]}\right]^{\mathrm{T}}\sim \mathcal{N}\left(\mathbf{0},\mathbf{Q}_{\boldsymbol{\beta}_{\mathrm{e}}[n]}\right)$ and the corresponding measurement noise covariance matrix is  $\mathbf{Q}_{\boldsymbol{\beta}_{\mathrm{e}}[n]} = \diag\{\sigma^2_{\tau _{\rm{e}}[n]},\sigma^2_{\nu _{\rm{e}}[n]},\sigma^2_{\sin \theta_{\mathrm{e}}[n]}, \sigma^2_{\sin \phi_{\mathrm{e}}[n]}\}$.
The non-linear function $\mathbf{g}_n\hspace{-2mm}: \mathbb{R}^{6 \times 1} \hspace{-1mm}\to \hspace{-1mm}\mathbb{R}^{4 \times 1}$ represents the measurement functions defined by \eqref{Eqn:MeasurementModel_I}.
We can observe that the measurement function $\mathbf{g}_n$ is time-varying as it depends on the state of the I-UAV, $\boldsymbol{\alpha}_b[n]$, and thus further on the navigation policy of the I-UAV in time slot $n$.
Moreover, the measurement covariance matrix $\mathbf{Q}_{\boldsymbol{\beta}_{\mathrm{e}}[n]}$ depends on the locations of both I-UAV and E-UAV as well as the AN precoding vector of the I-UAV in time slot $n$.
In other words, both the resource allocation and navigation designs affect the measurement performance.
Also, the measurement covariance matrix $\mathbf{Q}_{\boldsymbol{\beta}_{\mathrm{e}}[n]}$ is time-varying and can only be obtained online.
%

\subsubsection{E-UAV State Evolution Model}
In addition to the measurement model, a proper E-UAV state evolution model is required for realizing accurate tracking.
Assuming a constant velocity movement model \cite{FanLiuISAC}, the state evolution model of the E-UAV is given by
\vspace{-2mm}
\begin{equation}\label{Eqn:StateEvolution}
\boldsymbol{\alpha}_{\mathrm{e}} [n] =  \mathbf{F}\boldsymbol{\alpha}_{\mathrm{e}} [n-1] + \mathbf{v}_{\boldsymbol{\alpha}_{\mathrm{e}}},\vspace{-2mm}
\end{equation}
where $\mathbf{F}\in\mathbb{R}^{6 \times 6}$ is the state transition matrix, given by $\mathbf{F} = \left[ {\begin{array}{*{20}{c}}
	\mathbf{I}_3&\delta\mathbf{I}_3\\
	\mathbf{O}&\mathbf{I}_3
	\end{array}} \right]$, where $\mathbf{O}$ is a zero matrix.
Vector $\mathbf{v}_{\boldsymbol{\alpha}_{\mathrm{e}}} =  \left[v_{x_{\mathrm{e}}},v_{y_{\mathrm{e}}},v_{z_{\mathrm{e}}},v_{\dot x_{\mathrm{e}}},v_{\dot y_{\mathrm{e}}},v_{\dot z_{\mathrm{e}}}\right]^{\mathrm{T}} \in \mathbb{R}^{6 \times 1}$ denotes the state evolution noise with $v_{x_{\mathrm{e}}} \sim \mathcal{N}\left(0,\sigma^2_{x_{\mathrm{e}}}\right)$, $v_{y_{\mathrm{e}}} \sim \mathcal{N}\left(0,\sigma^2_{y_{\mathrm{e}}}\right)$, $v_{z_{\mathrm{e}}} \sim \mathcal{N}\left(0,\sigma^2_{z_{\mathrm{e}}}\right)$, $v_{\dot x_{\mathrm{e}}} \sim \mathcal{N}\left(0,\sigma^2_{ \dot x_{\mathrm{e}}}\right)$, $v_{\dot y_{\mathrm{e}}} \sim \mathcal{N}\left(0,\sigma^2_{\dot y_{\mathrm{e}}}\right)$, and $v_{\dot z_{\mathrm{e}}} \sim \mathcal{N}\left(0,\sigma^2_{\dot z_{\mathrm{e}}}\right)$.
Assuming mutually independent state evolution noises, we have $ \mathbf{v}_{\boldsymbol{\alpha}_{\mathrm{e}}}\sim \mathcal{N}\left(\mathbf{0},\mathbf{Q}_{\boldsymbol{\alpha}_{\mathrm{e}}}\right) $ with $\mathbf{Q}_{\boldsymbol{\alpha}_{\mathrm{e}}} = \diag\{\sigma^2_{x_{\mathrm{e}}},\sigma^2_{y_{\mathrm{e}}},\sigma^2_{z_{\mathrm{e}}},\sigma^2_{\dot x_{\mathrm{e}}},\sigma^2_{\dot y_{\mathrm{e}}},\sigma^2_{\dot z_{\mathrm{e}}}\}$, which can be acquired by the I-UAV based on long-term measurements. 

\subsubsection{E-UAV State Tracking via an EKF}
Due to the non-linear measurement model in \eqref{Eqn:MeautrementModel_II}, we adopt an extended Kalman filter (EKF) \cite{anderson2012optimal} to track the state of the E-UAV.
In time slot $n$, we assume that the I-UAV has an estimate of the E-UAV state in the previous time slot, $\hat{\boldsymbol{\alpha}}_{\mathrm{e}} [n\hspace{-1mm}-\hspace{-1mm}1]$, with corresponding covariance matrix $\mathbf{C}_{\mathrm{e}}[n\hspace{-1mm}-\hspace{-1mm}1] \hspace{-1mm} \in \hspace{-1mm} \mathbb{R}^{6 \times 6}$.
Then, the I-UAV predicts the state of the E-UAV in time slot $n$ as follows: 
\vspace{-2mm}
\begin{equation}\label{Eqn:StatePrediction}
\hat{\boldsymbol{\alpha}}_{\mathrm{e}} [n|n\hspace{-1mm}-\hspace{-1mm}1] \hspace{-1mm}=\hspace{-1mm} \mathbf{F}\hat{\boldsymbol{\alpha}}_{\mathrm{e}} [n\hspace{-1mm}-\hspace{-1mm}1] = [\hat{\mathbf{q}}^{\mathrm{T}}_{\mathrm{e}}[n|n\hspace{-1mm}-\hspace{-1mm}1], \hat{\dot{\mathbf{q}}}^{\mathrm{T}}_{\mathrm{e}}[n|n\hspace{-1mm}-\hspace{-1mm}1]]^{\mathrm{T}},\vspace{-2mm}
\end{equation}
where $\hat{\mathbf{q}}^{\mathrm{T}}_{\mathrm{e}}[n|n\hspace{-0.5mm}-\hspace{-0.5mm}1]$ and $\hat{\dot{\mathbf{q}}}^{\mathrm{T}}_{\mathrm{e}}[n|n\hspace{-0.5mm}-\hspace{-0.5mm}1]$ denote the predicted location and velocity of the E-UAV, respectively, and the corresponding prediction covariance matrix is given by
\vspace{-2mm}
\begin{equation}\label{Eqn:CovarianceMatrixPrediction}
\mathbf{C}_{\mathrm{e}}[n|n-1] = \mathbf{F}\mathbf{C}_{\mathrm{e}}[n-1]\mathbf{F}^{\mathrm{T}} + \mathbf{Q}_{\boldsymbol{\alpha}_{\mathrm{e}}}.\vspace{-2mm}
\end{equation}

Based on the predicted state of the E-UAV, the I-UAV designs its navigation strategy ${\bf{q}}_b\left[ n \right]$, predicts the wiretap channel $\hat{\mathbf{h}}_{\mathrm{be}}[n|n\hspace{-0.5mm}-\hspace{-0.5mm}1]$, and designs its resource allocation policy $\left({u}_k [\hspace{-0.25mm}n\hspace{-0.25mm}], \mathbf{w}_k [\hspace{-0.25mm}n\hspace{-0.25mm}], \mathbf{w}_{\mathrm{e}} [\hspace{-0.25mm}n\hspace{-0.25mm}]\right)$ for time slot $n$, as will be detailed in Section III.
Then, the I-UAV flies to its designed location, transmits data and AN according to the designed resource allocation strategy, and estimates the observable parameters in time slot $n$, $\hat{\boldsymbol{\beta}}_{\mathrm{e}} \hspace{-0.25mm}[\hspace{-0.25mm}n\hspace{-0.25mm}]$, which can subsequently be used to update the state of the E-UAV.

To update the state of the E-UAV in time slot $n$, the EKF employs a linearized measurement model around the predicted state, i.e.,
\vspace{-1mm}
\begin{equation}\label{Eqn:MeautrementModelLinearization}
\hat{\boldsymbol{\beta}}_{\mathrm{e}} \hspace{-0.25mm}[\hspace{-0.25mm}n\hspace{-0.25mm}] \hspace{-0.5mm}\approx\hspace{-0.5mm} \mathbf{g}_n\hspace{-1mm}\left(\hat{\boldsymbol{\alpha}}_{\mathrm{e}} [n|n\hspace{-1mm}-\hspace{-1mm}1]\right)\hspace{-0.5mm} +\hspace{-0.5mm} \mathbf{G}_n\hspace{-1mm}\left({\boldsymbol{\alpha}}_{\mathrm{e}}\hspace{-0.25mm} [\hspace{-0.25mm}n\hspace{-0.25mm}] \hspace{-1mm}-\hspace{-1mm} \hat{\boldsymbol{\alpha}}_{\mathrm{e}}\hspace{-0.25mm} [n|n\hspace{-1mm}-\hspace{-1mm}1]\right)\hspace{-0.5mm}+\hspace{-0.5mm} \mathbf{v}_{\boldsymbol{\beta}_{\mathrm{e}}}\hspace{-0.5mm}[\hspace{-0.25mm}n\hspace{-0.25mm}],\hspace{-1mm}\vspace{-1mm}
\end{equation}
where $\mathbf{G}_n \hspace{-1mm}\in\hspace{-1mm} \mathbb{R}^{4 \hspace{-0.25mm}\times\hspace{-0.25mm} 6}$ denotes the Jacobian matrix of $\mathbf{g}_n$ with respect to (w.r.t.) the predicted state of the E-UAV and is given by 
\vspace{-1mm}
\begin{equation}\label{Eqn:SensingMatrix}
{{\bf{G}}_n} = \left.\left[ {\begin{array}{*{20}{c}}
	{\frac{{\partial {\tau _{\rm{e}}}[n]}}{{\partial {x_{\rm{e}}}[n]}}}&{\frac{{\partial {\tau _{\rm{e}}}[n]}}{{\partial {y_{\rm{e}}}[n]}}}&{\frac{{\partial {\tau _{\rm{e}}}[n]}}{{\partial {z_{\rm{e}}}[n]}}}&0&0&0\\
	{\frac{{\partial {\nu _{\rm{e}}}[n]}}{{\partial {x_{\rm{e}}}[n]}}}&{\frac{{\partial {\nu _{\rm{e}}}[n]}}{{\partial {y_{\rm{e}}}[n]}}}&{\frac{{\partial {\nu _{\rm{e}}}[n]}}{{\partial {z_{\rm{e}}}[n]}}}&{\frac{{\partial {\nu _{\rm{e}}}[n]}}{{\partial {{\dot x}_{\rm{e}}}[n]}}}&{\frac{{\partial {\nu _{\rm{e}}}[n]}}{{\partial {{\dot y}_{\rm{e}}}[n]}}}&{\frac{{\partial {\nu _{\rm{e}}}[n]}}{{\partial {{\dot z}_{\rm{e}}}[n]}}}\\
	{\frac{{\partial \sin {\theta _{{\rm{e}}}}[n]}}{{\partial {x_{\rm{e}}}[n]}}}&{\frac{{\partial \sin {\theta _{{\rm{e}}}}[n]}}{{\partial {y_{\rm{e}}}[n]}}}&0&0&0&0\\
	{\frac{{\partial \sin {\phi _{{\rm{e}}}}[n]}}{{\partial {x_{\rm{e}}}[n]}}}&{\frac{{\partial \sin {\phi _{{\rm{e}}}}[n]}}{{\partial {y_{\rm{e}}}[n]}}}&{\frac{{\partial \sin {\phi _{{\rm{e}}}}[n]}}{{\partial {z_{\rm{e}}}[n]}}}&0&0&0
	\end{array}} \right]\right|_{\hat{\boldsymbol{\alpha}}_{\mathrm{e}} [n|n-1]}.\vspace{-1mm}
\end{equation}
Note that the entries of ${{\bf{G}}_n}$ can be straightforwardly derived from \eqref{Eqn:MeasurementModel_I} by replacing ${\boldsymbol{\alpha}}_{\mathrm{e}} [n]$ with $\hat{\boldsymbol{\alpha}}_{\mathrm{e}} [n|n-1]$, which is omitted here due to the page limitation.
Now, the state of the E-UAV in time slot $n$ is estimated as
\vspace{-2mm}
\begin{equation}\label{Eqn:StateTracking}
\hat{{\boldsymbol{\alpha}}}_{\mathrm{e}} [n] = \hat{\boldsymbol{\alpha}}_{\mathrm{e}} [n|n-1] + \mathbf{K}_n\left(\hat{\boldsymbol{\beta}}_{\mathrm{e}} [n] - \mathbf{g}_n\left(\hat{\boldsymbol{\alpha}}_{\mathrm{e}} [n|n-1]\right)\right)\vspace{-2mm}
\end{equation}
and the corresponding posterior covariance matrix is given by
\vspace{-2mm}
\begin{equation}\label{Eqn:CovPosteriorMSE}
\mathbf{C}_{\mathrm{e}}[n] = \left(\mathbf{I}_6 - \mathbf{K}_n{{\bf{G}}_n}\right)\mathbf{C}_{\mathrm{e}}[n|n-1] \in \mathbb{R}^{6 \times 6},\vspace{-2mm}
\end{equation}
where the Kalman gain matrix $\mathbf{K}_n \in \mathbb{R}^{6 \times 4}$ is given by
\vspace{-2mm}
\begin{equation}\label{Eqn:KalmanGain}
\mathbf{K}_n = \mathbf{C}_{\mathrm{e}}[n|n\hspace{-1mm}-\hspace{-1mm}1]{{\bf{G}}^{\mathrm{T}}_n}\hspace{-1mm}\left({{\bf{G}}_n}\mathbf{C}_{\mathrm{e}}[n|n\hspace{-1mm}-\hspace{-1mm}1]{{\bf{G}}^{\mathrm{T}}_n}\hspace{-1mm}+\hspace{-1mm}\hat{\mathbf{Q}}_{\boldsymbol{\beta}_{\mathrm{e}}[n]}\right)^{-1}.\vspace{-2mm}
\end{equation}

Note that $\hat{{\boldsymbol{\alpha}}}_{\mathrm{e}}[n]$ and $\mathbf{C}_{\mathrm{e}}[n]$ are the posterior mean and covariance matrix of the hidden state variable ${{\boldsymbol{\alpha}}}_{\mathrm{e}} [n]$ in time slot $n$ determined by combining the information obtained from state prediction $\hat{\boldsymbol{\alpha}}_{\mathrm{e}} [n|n-1]$ and the new measurement $\hat{\boldsymbol{\beta}}_{\mathrm{e}} [n]$.
In fact, the trace of $\mathbf{C}_{\mathrm{e}}[n]$ characterizes the posterior mean squared error (MSE) for tracking the state of the E-UAV, which needs to be considered for resource allocation design.
Moreover, as the ground truth of $d_{\mathrm{e}}[n]$ needed in \eqref{Eqn:MFsnr} is unknown to the EKF, the measurement covariance matrix 
$\hat{\mathbf{Q}}_{\boldsymbol{\beta}_{\mathrm{e}}[\hspace{-0.25mm}n\hspace{-0.25mm}]}$ in \eqref{Eqn:KalmanGain} is estimated based on new measurements, i.e., $\hat{\mathbf{Q}}_{\boldsymbol{\beta}_{\mathrm{e}}[\hspace{-0.25mm}n\hspace{-0.25mm}]} = \mathbf{Q}_{\boldsymbol{\beta}_{\mathrm{e}}[n]}\left|_{\hat{\boldsymbol{\beta}}_{\mathrm{e}} [n]}\right..$
Substituting \eqref{Eqn:KalmanGain} into \eqref{Eqn:CovPosteriorMSE} and employing the matrix inversion lemma yields
\vspace{-2mm}
\begin{equation}\label{Eqn:CovPosteriorMSEInverse}
{\mathbf{C}}_{\mathrm{e}}[n] = \left(\mathbf{C}^{-1}_{\mathrm{e}}[n|n-1] + {{\bf{G}}^{\mathrm{T}}_n}\hat{\mathbf{Q}}^{-1}_{\boldsymbol{\beta}_{\mathrm{e}}[n]}{{\bf{G}}_n}\right)^{-1},\vspace{-2mm}
\end{equation}
from which ${\mathbf{C}}^{-1}_{\mathrm{e}}[n]$ can be easily obtained and will be exploited for resource allocation design in Section III-C.

\vspace{-2mm}
\section{Proposed Online Navigation and Resource Allocation Design}
In this section, based on the proposed ISNC framework, the online navigation and resource allocation designs for the I-UAV are developed by exploiting the state estimates of the E-UAV obtained in the previous time slot.

\vspace{-4mm}
\subsection{Proposed Distance-based Online Navigation Policy}
\begin{figure}[t]
	\vspace{-5mm}
	\center{\includegraphics[width=2.8in]{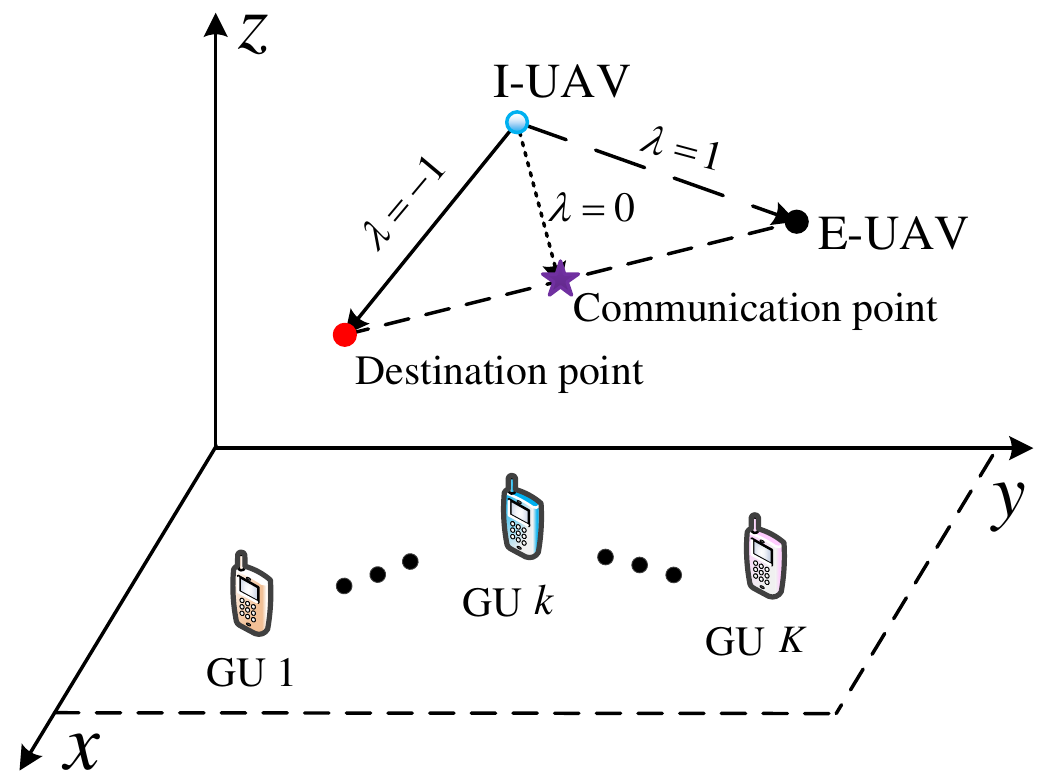}}\vspace{-7mm}
	\caption{Proposed distance-based navigation policy for the considered system.}
	\label{fig:NavModel}
	\vspace{-10mm}
\end{figure}

As a causal system, the navigation policy of the I-UAV in the current time slot affects the sensing and communication performance in the following time slots.
However, predicting the future sensing and communication performance based on the currently adopted navigation policy is very challenging, and thus, a globally optimal navigation design for determining the best possible trade-off between sensing and communication is intractable.
To facilitate fast online navigation design, we propose a low-complexity distance-based suboptimal navigation policy, as illustrated in Fig. \ref{fig:NavModel}.
In particular, let us define a communication point whose horizontal location is determined by the centriod of all the GUs and whose altitude is fixed as the minimum possible altitude of the I-UAV $z_{\mathrm{min}}$, i.e., $\mathbf{q}_{\mathrm{Com}} = \frac{1}{K}\sum_{k=1}^{K}\mathbf{q}_k + \left[0,0,z_{\mathrm{min}}\right]^{\mathrm{T}}$.
Then, we define a desired destination point on the line between the communication point and the predicted location of the E-UAV in time slot $n$.
The location of the desired destination point is given by
\vspace{-1mm}
\begin{equation}\label{Eqn:Navigation_Destination}
	\mathbf{q}_{\mathrm{Des}}[n] = (1-\lambda)\mathbf{q}_{\mathrm{Com}} + \lambda\hat{\mathbf{q}}_{\mathrm{e}}[n|n\hspace{-1mm}-\hspace{-1mm}1],\vspace{-1mm}
\end{equation}
where $-1 < \lambda < 1$ is a parameter to control the trade-off between sensing, jamming, and communications.
Increasing $\lambda$ guides the I-UAV towards the predicted location of the E-UAV,  $\hat{\mathbf{q}}_{\mathrm{e}}[n|n\hspace{-1mm}-\hspace{-1mm}1]$, in time slot $n$, which is beneficial for jamming and sensing.
On the other hand, decreasing $\lambda$ guides the I-UAV away from the predicted location of the E-UAV,  $\hat{\mathbf{q}}_{\mathrm{e}}[n|n\hspace{-1mm}-\hspace{-1mm}1]$, in time slot $n$, which might be beneficial for secure communications.
%

In time slot $n$, the I-UAV navigator needs to optimize its acceleration to minimize its distance w.r.t. the desired destination point, which leads to the following optimization problem
\vspace{-2mm}
\begin{align}\label{Eqn:NavProblemFormulationII}
	\underset{\ddot {\bf{q}}_{\mathrm{b}}\left[ n\right]}{\mathrm{minimize}}\,\; & \left\| \mathbf{q}_{\mathrm{Des}}[n] - {\bf{q}}_{\mathrm{b}}\left[ n \right] \right\|
	\\[-1mm]
	\notag\mbox{s.t.}\;\;
	\mbox{{C1:}}\; & \left\|\dot {\bf{q}}_{\mathrm{b}}\left[ n \right] \right\| \le V_{\mathrm{max}}, \notag\\[-1mm]
	\mbox{{C2:}}\; & \left|{\ddot{x}}_{\mathrm{b}}[n] \right| \hspace{-1mm}\le\hspace{-1mm} A^x_{\mathrm{cc}}, \left|{\ddot{y}}_{\mathrm{b}}[n] \right|\hspace{-1mm} \le\hspace{-1mm}  A^y_{\mathrm{cc}}, \left|{\ddot{z}}_{\mathrm{b}}[n] \right| \hspace{-1mm}\le\hspace{-1mm} A^z_{\mathrm{cc}}, \notag\\[-1mm]
	\mbox{{C3:}}\;& \mathbf{q}_{\mathrm{min}} \le {\mathbf{q}}_{\mathrm{b}}\left[ n+l \right] \le \mathbf{q}_{\mathrm{max}}, \forall l=0,\ldots,L-1,\notag\\[-1mm]
	\mbox{{C4:}}\;& \max\left\{| \hat{x}_{{\mathrm{e}}}[n+l|n \hspace{-1mm}-\hspace{-1mm}1] - {x}_{\mathrm{b}}\left[ n+l \right] |,| \hat{y}_{{\mathrm{e}}}[n+l|n \hspace{-1mm}-\hspace{-1mm}1] - {y}_{\mathrm{b}}\left[ n+l \right] |,\right.\notag\\[-1mm]
	&\left.| \hat{z}_{{\mathrm{e}}}[n+l|n \hspace{-1mm}-\hspace{-1mm}1] - {z}_{\mathrm{b}}\left[ n+l \right] |\right\} \ge d_{\mathrm{min}}, \forall l=0,\ldots,L-1,\notag
\end{align}
\vspace{-8mm}\par\noindent
where optimization variables $\ddot {\bf{q}}_{\mathrm{b}}\left[ n\right] = \left[{\ddot{x}}_{\mathrm{b}}[n], {\ddot{y}}_{\mathrm{b}}[n],{\ddot{z}}_{\mathrm{b}}[n]\right]^{\mathrm{T}}$ are the 3D acceleration variables of the I-UAV in time slot $n$.
Constant $V_{\mathrm{max}}$ in C1 denotes the maximum flying velocity of the I-UAV.
In C2, $A^x_{\mathrm{cc}}$, $A^y_{\mathrm{cc}}$, and $A^z_{\mathrm{cc}}$ denote the maximum acceleration in each dimension.
C3 limits the I-UAV to fly within the considered 3D space $\left[\mathbf{q}_{\mathrm{min}},\mathbf{q}_{\mathrm{max}}\right]$ in the subsequent $L$ time slots, where $\mathbf{q}_{\mathrm{min}} = \left[x_{\mathrm{min}},y_{\mathrm{min}},z_{\mathrm{min}}\right]^{\mathrm{T}}$ and $\mathbf{q}_{\mathrm{max}} = \left[x_{\mathrm{max}},y_{\mathrm{max}},z_{\mathrm{max}}\right]^{\mathrm{T}}$.
C4 is a collision avoidance constraint for subsequent $L$ time slots, where $d_{\mathrm{min}}$ is the allowable safe distance between I-UAV and E-UAV.

By assuming constant acceleration in the subsequent $L$ time slots, the velocity and location of the I-UAV in time slot $n+l$, $\forall l=0,\ldots,L-1$, in \eqref{Eqn:NavProblemFormulationII} are predicted as 
\vspace{-2mm}
\begin{align}
	\dot{\mathbf{q}}_{\mathrm{b}}\left[ n+l \right] &= \dot{\mathbf{q}}_{\mathrm{b}}\left[ n \hspace{-1mm}-\hspace{-1mm}1 \right] + \ddot {\mathbf{q}}_{\mathrm{b}}\left[ n \right] (l+1)\delta \hspace{5mm}\text{and}\notag\\[-1mm]
	{\mathbf{q}}_{\mathrm{b}}\left[ n+l \right] &= {\mathbf{q}}_{\mathrm{b}}\left[ n \hspace{-1mm}-\hspace{-1mm}1 \right] + \dot{\mathbf{q}}_{\mathrm{b}}\left[ n \hspace{-1mm}-\hspace{-1mm}1 \right] (l+1)\delta + \frac{1}{2}\ddot{\mathbf{q}}_{\mathrm{b}}\left[ n \right] (l+1)^2\delta^2,
\end{align}
\vspace{-9mm}\par\noindent
respectively. 
Furthermore, $\hat{\mathbf{q}}_{\mathrm{e}}[n+l|n-1] = \left[\hat{x}_{{\mathrm{e}}}[n+l|n \hspace{-1mm}-\hspace{-1mm}1],\hat{y}_{{\mathrm{e}}}[n+l|n \hspace{-1mm}-\hspace{-1mm}1],\hat{z}_{{\mathrm{e}}}[n+l|n \hspace{-1mm}-\hspace{-1mm}1]\right]^{\mathrm{T}}$ in C4 is the predicted location of the E-UAV in time slot $n+l$, which can be obtained from
\vspace{-1mm}
\begin{equation}\label{Eqn:StatePredictionMulti-Step}
	\hat{\boldsymbol{\alpha}}_{\mathrm{e}} [n+l|n\hspace{-1mm}-\hspace{-1mm}1] \hspace{-1mm}=\hspace{-1mm} \mathbf{F}^{l+1}\hat{\boldsymbol{\alpha}}_{\mathrm{e}} [n\hspace{-1mm}-\hspace{-1mm}1] = [\hat{\mathbf{q}}^{\mathrm{T}}_{\mathrm{e}}[n+l|n\hspace{-1mm}-\hspace{-1mm}1], \hat{\dot{\mathbf{q}}}^{\mathrm{T}}_{\mathrm{e}}[n+l|n\hspace{-1mm}-\hspace{-1mm}1]]^{\mathrm{T}}.\vspace{-1mm}
\end{equation}

Constraint C4 can be equivalently replaced by the following four constraints by introducing auxiliary optimization variables $d^l_{\mathrm{C4}}$
\vspace{-1mm}
\begin{align}
	\mbox{{C4a:}}\;& | \hat{x}_{{\mathrm{e}}}[n+l|n \hspace{-1mm}-\hspace{-1mm}1] - {x}_{\mathrm{b}}\left[ n+l \right] | \le d^l_{\mathrm{C4}}, \;\;
	\mbox{{C4b:}}\; | \hat{y}_{{\mathrm{e}}}[n+l|n \hspace{-1mm}-\hspace{-1mm}1] - {y}_{\mathrm{b}}\left[ n+l \right] | \le d^l_{\mathrm{C4}}, \notag\\[-1mm]
	\mbox{{C4c:}}\;& | \hat{z}_{{\mathrm{e}}}[n+l|n \hspace{-1mm}-\hspace{-1mm}1] - {z}_{\mathrm{b}}\left[ n+l \right] | \le d^l_{\mathrm{C4}}, \;\;\;
	\mbox{{C4d:}}\; d^l_{\mathrm{C4}} \ge d_{\mathrm{min}},\forall l=0,\ldots,L-1.
\end{align}
\vspace{-8mm}\par\noindent
Now, the problem formulated in \eqref{Eqn:NavProblemFormulationII} is convex and can be solved efficiently by off-the-shelf numerical convex solvers, such as CVX\cite{book:convex}.

Note that imposing constraints C3 and C4 for the subsequent $L$ time slots helps in coping with the inherent momentum of the I-UAV.
In particular, when $L=1$, i.e., for a single-step navigation design, the I-UAV may sometimes not be able to avoid collisions or leaving the considered 3D space due to its inertia of motion causing the violation of C3 or C4, respectively.
For instance, if the I-UAV is close to the boundary of the considered 3D space in a given time slot and has a large velocity towards the boundary, none of the feasible acceleration actions satisfying C2 may be able to prevent the I-UAV from leaving the considered 3D space, i.e., the violation of C3 may be unavoidable, which results in an infeasible problem.
However, when $L>1$,
if the I-UAV approaches the boundary or E-UAV, the proposed multi-step navigation design identifies potential future infeasibilities ahead of time and takes corresponding action to avoid a collision or leaving the considered 3D space.
The larger $L$ is chosen, the more conservative the navigation design.

If the problem in \eqref{Eqn:NavProblemFormulationII} is feasible, the I-UAV flies according to the optimized acceleration in the next time slot.
On the other hand, if the problem in \eqref{Eqn:NavProblemFormulationII} is infeasible, a compromise navigation policy should be adopted to minimize the distance between the I-UAV and the desired destination while penalizing possible collisions or leaving the considered 3D space.
For the time slots when the problem in \eqref{Eqn:NavProblemFormulationII} is infeasible, we propose to adjust the navigation strategy by solving the following optimization problem:
\vspace{-2mm}
\begin{align}\label{Eqn:NavProblemFormulationIII}
	\underset{\ddot {\bf{q}}_{\mathrm{b}}\left[ n\right],d^l_{\mathrm{C4}},d^l_{\mathrm{C3a}},d^l_{\mathrm{C3b}}}{\mathrm{minimize}}\,\; & \left\| \mathbf{q}_{\mathrm{Des}}[n] - {\bf{q}}_{\mathrm{b}}\left[ n \right] \right\| + \omega_{\mathrm{C3}}\max\{ d^l_{\mathrm{C3a}},0\} \notag\\[-3mm]
	&+ \omega_{\mathrm{C3}}\max\{ d^l_{\mathrm{C3b}},0\} + \omega_{\mathrm{C4}}\max\{d_{\mathrm{min}} - d^l_{\mathrm{C4}},0\}
	\\[-1mm]
	\notag\mbox{s.t.}\;\;
	&\mbox{{C1,}}\; \mbox{{C2,}}\; \mbox{{C4a,}} \; \mbox{{C4b,}}\; \mbox{{C4c,}} \notag\\[-2mm]
	&\mbox{{C3a:}}\; \mathbf{q}_{\mathrm{min}} - {\mathbf{q}}_{\mathrm{b}}\left[ n+l \right] \le d^l_{\mathrm{C3a}}, \forall l=0,\ldots,L-1, \notag\\[-1mm]
	&\mbox{{C3b:}}\; {\mathbf{q}}_{\mathrm{b}}\left[ n+l \right] - \mathbf{q}_{\mathrm{max}} \le d^l_{\mathrm{C3b}}, \forall l=0,\ldots,L-1, \notag
\end{align}
\vspace{-9mm}\par\noindent
where $d^l_{\mathrm{C3a}},d^l_{\mathrm{C3b}},d^l_{\mathrm{C4}}\ge 0$ are auxiliary optimization variables, which are used to penalize the violation of constraints C3 and C4 in \eqref{Eqn:NavProblemFormulationII}, and $\omega_{\mathrm{C3}}, \omega_{\mathrm{C4}} \ge 0$ are the corresponding penalty scaling factors. 
When C3 in \eqref{Eqn:NavProblemFormulationII} cannot be satisfied due to the momentum of the E-UAV, $d^l_{\mathrm{C3a}}$ or $d^l_{\mathrm{C3b}}$ are positive and this causes a penalty, otherwise, $d^l_{\mathrm{C3a}} = d^l_{\mathrm{C3b}} = 0$.
Also, when C4 in \eqref{Eqn:NavProblemFormulationII} is violated, $d^l_{\mathrm{C4}}$ is positive, which causes a corresponding penalty, otherwise, $d^l_{\mathrm{C4}} = 0$.

In summary, the proposed suboptimal navigation design first solves \eqref{Eqn:NavProblemFormulationII} to obtain an effective navigation policy to minimize the distance between the I-UAV and the desired destination point. 
If \eqref{Eqn:NavProblemFormulationII} is infeasible, we solve \eqref{Eqn:NavProblemFormulationIII} to obtain a compromise navigation policy.
Compared to the single-step navigation design with $L=1$, the proposed multi-step navigation design with $L>1$ can predictively identify whether the I-UAV is approaching the service area boundary or the E-UAV and cause the I-UAV to abandon its current moving direction.

\vspace{-4mm}
\subsection{Wiretap Channel Prediction}
For efficient jamming, the I-UAV has to acquire the effective wiretap channel information between I-UAV and E-UAV for resource allocation design.
Based on the predicted state of the E-UAV in \eqref{Eqn:StatePrediction} and the  navigation policy designed in Section III-A, the effective wiretap channel between I-UAV and E-UAV in time slot $n$ can be predicted as follows
\vspace{-1mm}
\begin{equation}\label{Eqn:PredictedChannel}
\hat{\mathbf{h}}_{\mathrm{be}}[n|n\hspace{-1mm}-\hspace{-1mm}1] \hspace{-1mm}=\hspace{-1mm} \frac{\beta_0}{\hat{d}_{\mathrm{e}}\hspace{-0.25mm}[\hspace{-0.25mm}n|n\hspace{-1mm}-\hspace{-1mm}1\hspace{-0.25mm}]}\mathbf{a}_{{M^{\mathrm{tx}}_{\mathrm{b}}},{M^{\mathrm{ty}}_{\mathrm{b}}}}\hspace{-1mm}\left(\hspace{-0.5mm}\hat{\theta}_{\mathrm{e}}\hspace{-0.25mm}[\hspace{-0.25mm}n|n\hspace{-1mm}-\hspace{-1mm}1\hspace{-0.25mm}], \hat{\phi}_{\mathrm{e}}\hspace{-0.25mm}[\hspace{-0.25mm}n|n\hspace{-1mm}-\hspace{-1mm}1\hspace{-0.25mm}]\hspace{-0.5mm}\right),\vspace{-1mm}
\end{equation}
where $\hat{d}_{\mathrm{e}}[n|n\hspace{-1mm}-\hspace{-1mm}1]$, $\hat{\theta}_{\mathrm{e}}[n|n\hspace{-1mm}-\hspace{-1mm}1]$, and $\hat{\phi}_{\mathrm{e}}[n|n\hspace{-1mm}-\hspace{-1mm}1]$ are the predicted distance, azimuth AOD, and elevation AOD, respectively, which are given by
\vspace{-2mm}
\begin{align}\label{Eqn:PredictedChannel_Para}
	\hat{d}_{\mathrm{e}}[n|n\hspace{-1mm}-\hspace{-1mm}1] &= 
	\left\| \hat{\mathbf{q}}_{\mathrm{e}}[n|n\hspace{-1mm}-\hspace{-1mm}1] - {\bf{q}}_{{\mathrm{b}}}\left[ n \right] \right\|\notag\\[-1mm]
    \hat{\theta}_{\mathrm{e}}[n|n\hspace{-1mm}-\hspace{-1mm}1] &\hspace{-1mm}=\hspace{-1mm} \arcsin \left(\frac{x_{\mathrm{e}}[n|n\hspace{-1mm}-\hspace{-1mm}1] \hspace{-1mm}-\hspace{-1mm} x_{\mathrm{b}}[n]}{\sqrt{\left|x_{\mathrm{e}}[n|n\hspace{-1mm}-\hspace{-1mm}1] \hspace{-1mm}-\hspace{-1mm} x_{\mathrm{b}}[n]\right|^2\hspace{-1mm}+\hspace{-1mm}\left|y_{\mathrm{e}}[n|n\hspace{-1mm}-\hspace{-1mm}1] \hspace{-1mm}-\hspace{-1mm} y_{\mathrm{b}}[n]\right|^2}}\right),\;\text{and}\notag\\[-1mm]
	\hat{\phi}_{\mathrm{e}}[n|n\hspace{-1mm}-\hspace{-1mm}1] &\hspace{-1mm}=\hspace{-1mm} \arcsin \left(\frac{z_{\mathrm{e}}[n|n\hspace{-1mm}-\hspace{-1mm}1] - z_{\mathrm{b}}[n]}{\|\mathbf{q}_{\mathrm{e}} [n|n\hspace{-1mm}-\hspace{-1mm}1] - \mathbf{q}_{{\mathrm{b}}}[n]\|}\right),
\end{align}
\vspace{-6mm}\par\noindent
respectively.
%
%
Because of the E-UAV state prediction uncertainty characterized by \eqref{Eqn:CovarianceMatrixPrediction}, the channel prediction in \eqref{Eqn:PredictedChannel} is inevitably imperfect, which has to be taken into account for resource allocation design.
Analyzing the prediction error variance is a challenging task due to complicated non-linear mapping between $\hat{\boldsymbol{\alpha}}_{\mathrm{e}} [n|n\hspace{-1mm}-\hspace{-1mm}1]$ and $\hat{\mathbf{h}}_{\mathrm{be}}[n|n\hspace{-1mm}-\hspace{-1mm}1]$ in \eqref{Eqn:PredictedChannel} and \eqref{Eqn:PredictedChannel_Para}.

\begin{figure}[t]
	\begin{minipage}{.55\textwidth}
		\vspace{2mm}
		\center{\includegraphics[width=3.55in]{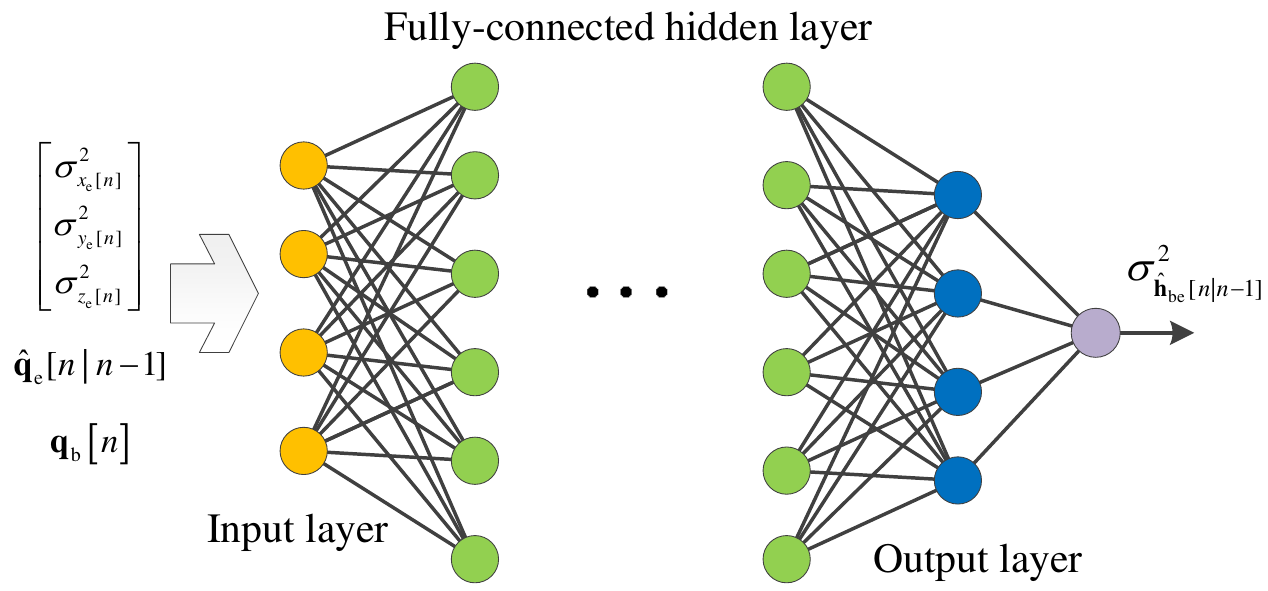}}
		\caption{Proposed DNN-based channel prediction error bound estimation.}
		\label{fig:ChannelPredictModel}
		\vspace{-7mm}
	\end{minipage}
	\begin{minipage}{.4\textwidth}
		\vspace{-3mm}
		\center{\includegraphics[width=3in]{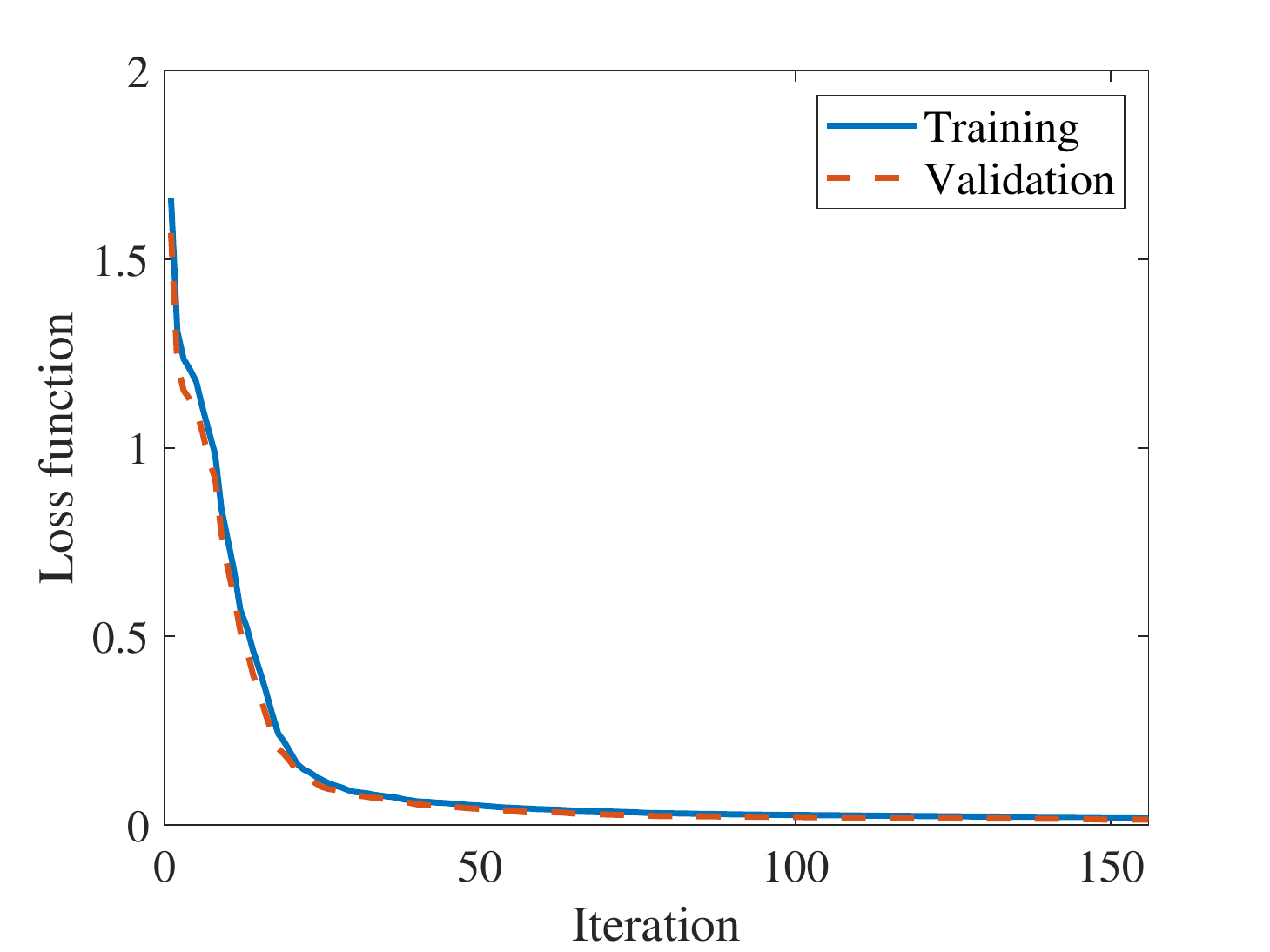}}\vspace{-7mm}
		\caption{Training process of the proposed DNN-based channel prediction error bound estimation.}
		\label{fig:ChannelPredictModel_Training}
		\vspace{-10mm}
	\end{minipage}
\end{figure}

To facilitate robust resource allocation design, we assume that the effective wiretap channel prediction error, $\Delta \hat{\mathbf{h}}_{\mathrm{be}}[n|n\hspace{-1mm}-\hspace{-1mm}1]$, can be captured by a bounded error model, i.e., $\left\|\Delta \hat{\mathbf{h}}_{\mathrm{be}}[n|n\hspace{-1mm}-\hspace{-1mm}1]\right\|^2 \hspace{-1mm}\le\hspace{-1mm} \sigma^2_{\hat{\mathbf{h}}_{\mathrm{be}}[n|n\hspace{-0.25mm}-\hspace{-0.25mm}1]}$, where $\sigma^2_{\hat{\mathbf{h}}_{\mathrm{be}}[n|n\hspace{-0.25mm}-\hspace{-0.25mm}1]}$ is estimated by a deep neural network (DNN), as illustrated in Fig. \ref{fig:ChannelPredictModel}.
The input of the DNN comprises the predicted location of the E-UAV, the corresponding prediction error variance, and the location of the I-UAV, while the output of the DNN is the bound on the channel prediction error.

During the offline training phase, in the $i$-th training episode, we uniformly and randomly generate the locations of the E-UAV and the I-UAV in the considered 3D space as $\mathbf{q}^{i}_{\mathrm{b}}[n]$ and $\mathbf{q}^{i}_{\mathrm{e}}[n]$, respectively.
Also, we assume that the prediction error variances $({\sigma _{i,{x_\mathrm{e}}[n]}^2},{\sigma _{i,{y_\mathrm{e}}[n]}^2},{\sigma _{i,{z_\mathrm{e}}[n]}^2})$ are sampled within $\left[0,10\right]$ uniformly and randomly.
For each episode, we generate $J$ imperfect predicted locations of the E-UAV according to the following model:
\vspace{-1mm}
\begin{equation}\label{Eqn:TrainingData}
	\hat{\mathbf{q}}^{ij}_{\mathrm{e}}[n|n\hspace{-1mm}-\hspace{-1mm}1] = \mathbf{q}^{i}_{\mathrm{e}}[n] + \mathbf{v}^{ij}_{\mathbf{q}_{\mathrm{e}}}[n], j = 1,\ldots, J,\vspace{-1mm}
\end{equation}
where $\mathbf{v}^{ij}_{\mathbf{q}_{\mathrm{e}}}[n]\sim \mathcal{N}\left(\mathbf{0},\diag({\sigma _{i,{x_\mathrm{e}}[n]}^2},{\sigma _{i,{y_\mathrm{e}}[n]}^2},{\sigma _{i,{z_\mathrm{e}}[n]}^2})\right)$.
Substituting $\mathbf{q}^{i}_{\mathrm{e}}[n]$ and $\hat{\mathbf{q}}^{ij}_{\mathrm{e}}[n|n-1]$ into \eqref{Eqn:EffectiveChannel} and \eqref{Eqn:PredictedChannel}, we can obtain the actual wiretap channel ${\mathbf{h}}^{i}_{\mathrm{be}}[n]$ and the corresponding predicted wiretap channel $\hat{\mathbf{h}}^{ij}_{\mathrm{be}}[n|n-1]$, respectively.
Now, we obtain the labeling for the channel prediction error bound for the DNN input $\left({\sigma _{i,{x_\mathrm{e}}[n]}^2},{\sigma _{i,{y_\mathrm{e}}[n]}^2},{\sigma _{i,{z_\mathrm{e}}[n]}^2},\mathbf{q}^{i}_{\mathrm{e}}[n],{\bf{q}}^{i}_b\left[ n \right]\right)$ in the $i$-th episode as 
\vspace{-1mm}
\begin{equation}
	\tilde{\sigma}^2_{i,\hat{\mathbf{h}}_{\mathrm{be}}[n|n\hspace{-0.25mm}-\hspace{-0.25mm}1]} = \max_{j} \left\|{\mathbf{h}}^{i}_{\mathrm{be}}[n] - \hat{\mathbf{h}}^{ij}_{\mathrm{be}}[n|n\hspace{-1mm}-\hspace{-1mm}1] \right\|^2.\vspace{-1mm}
\end{equation}
Define the loss function for training the DNN as follows
\vspace{-1mm}
\begin{equation}
\mathcal{L} = \frac{1}{2}\left|\tilde{\sigma}^2_{i,\hat{\mathbf{h}}_{\mathrm{be}}[n|n\hspace{-0.25mm}-\hspace{-0.25mm}1]} - {\sigma}^2_{i,\hat{\mathbf{h}}_{\mathrm{be}}[n|n\hspace{-0.25mm}-\hspace{-0.25mm}1]}\right|^2,\vspace{-1mm}
\end{equation}
where ${\sigma}^2_{i,\hat{\mathbf{h}}_{\mathrm{be}}[n|n\hspace{-0.25mm}-\hspace{-0.25mm}1]}$ is the DNN output in the $i$-th training episode.
The weights and biases of the DNN can be updated by exploiting a gradient descent-based back propagation learning rule, such as Adam \cite{Adam2015}.
An exemplary offline training process is depicted in Fig. \ref{fig:ChannelPredictModel_Training}.
The system parameters adopted in the training process are specified in Table \ref{simulation_setting} in Section IV.
A three-layer feedforward fully-connected neural network with layer sizes of 20, 100, and 20, respectively, is adopted, where each layer is followed by a rectified linear unit (ReLU) activation function except for the final fully connected layer.
For generating training samples, we utilize 10000 episodes and $J=1000$ for each episode.

During the online test stage, the predicted location of the E-UAV in time slot $n$ is determined according to \eqref{Eqn:StatePrediction} and corresponding prediction variances are obtained according to \eqref{Eqn:CovarianceMatrixPrediction}, i.e.,
${\sigma _{{x_\mathrm{e}[n]}}^2} = {{{\left\{ {{{\bf{C}}_{\rm{e}}}[n|n - 1]} \right\}}_{11}}}$, ${\sigma _{{y_\mathrm{e}[n]}}^2} = {{{\left\{ {{{\bf{C}}_{\rm{e}}}[n|n - 1]} \right\}}_{22}}}$, and ${\sigma _{{z_\mathrm{e}[n]}}^2} = {{{\left\{ {{{\bf{C}}_{\rm{e}}}[n|n - 1]} \right\}}_{33}}}$.
The location of the I-UAV is found with the proposed online navigation design.
Then, the trained DNN provides an estimate for the channel prediction error bound, which is subsequently used for robust resource allocation design.

\vspace{-4mm}
\subsection{Proposed Robust Resource Allocation Design}
Based on the predicted effective wiretap channel in \eqref{Eqn:PredictedChannel} and the designed navigation policy in Section III-A, in time slot $n$, the I-UAV needs to determine its resource allocation strategy to achieve secure communication and accurate sensing. 
In fact, communication and sensing have to be optimized simultaneously since the former is the goal of the I-UAV and the latter is a prerequisite for efficient jamming in the following time slots. 
In time slot $n$, the objective of the proposed robust resource allocation design is to maximize the number of securely served GUs and to simultaneously minimize the tracking MSE, which leads to the following optimization problem:
\begin{align}\label{Eqn:ProblemFormulation}
\underset{{u}_k [\hspace{-0.25mm}n\hspace{-0.25mm}], \mathbf{w}_k [\hspace{-0.25mm}n\hspace{-0.25mm}], \mathbf{w}_{\mathrm{e}} [\hspace{-0.25mm}n\hspace{-0.25mm}]}{\mathrm{maximize}}\,\; & \frac{1}{1+\omega}\frac{\sum\nolimits_{k=1}^{K} u_k[n]}{K} - \frac{\omega}{1+\omega} \frac{\mathrm{Tr}\{{\mathbf{C}}_{\mathrm{e}}[n]\}}{\mathrm{MSE}_{\mathrm{max}}}
\\[-1mm]
\notag\mbox{s.t.}\;\;
\mbox{{C1:}}\; &\sum\nolimits_{k = 1}^{K} {u}_k [n] \left\|\mathbf{w}_k [n]\right\|^2 + \left\|\mathbf{w}_{\mathrm{e}} [n]\right\|^2 \le p_{\mathrm{max}},\notag\\[-1mm]
\mbox{{C2:}}\; & {{R}_k}[n] \ge {u}_k [n] {R_{\mathrm{min}}}, \forall k \in \mathcal{K},\notag\\[-1mm]
\mbox{{C3:}}\; & 
\underset{\left\|\Delta \hat{\mathbf{h}}_{\mathrm{be}}[n|n-1]\right\|^2 \le \sigma^2_{\hat{\mathbf{h}}_{\mathrm{be}}[n|n\hspace{-0.25mm}-\hspace{-0.25mm}1]}}{\mathrm{max}}
{{R}^k_{\rm{e}}}[n|n-1] \le {u}_k [n] {R_{\mathrm{Leakage}}}, \forall k \in \mathcal{K}.\notag 
\end{align}
\vspace{-4mm}\par\noindent
In the objective function, $\sum\nolimits_{k=1}^{K} u_k[n]$ is the total number of GUs that can be served securely in time slot $n$, which is normalized by $K$ and characterizes the secure communication performance.
Besides, $\mathrm{Tr}\{{\mathbf{C}}_{\mathrm{e}}[n]\}$ is the tracking MSE in time slot $n$ obtained from \eqref{Eqn:CovPosteriorMSEInverse}, which is normalized by a pre-fixed constant $\mathrm{MSE}_{\mathrm{max}}$ and characterizes the sensing performance.
The non-negative parameter $\omega \ge 0$ is used to control the trade-off between communication and sensing performance.
Furthermore, C1 imposes a transmit power constraint, where $p_{\mathrm{max}}$ denotes the maximum transmit power of the I-UAV.
C2 is the communication rate constraint for the selected GUs, where ${R_{\mathrm{min}}}$ is the minimum required data rate of each served GU.
In C3, ${{R}^k_{\rm{e}}}[n|n-1]$ is the predicted leakage rate obtained by substituting the predicted effective wiretap channel \eqref{Eqn:PredictedChannel} into \eqref{Eqn:EavesdroppingUAVRate}, as the actual wiretap channel is not known at the time of resource allocation.
Also, ${R_{\mathrm{Leakage}}}$ in C3 is the maximum tolerable leakage rate associated with the selected GUs in the presence of wiretap channel prediction uncertainty.

Introducing an auxiliary tracking MSE variable $\xi[n]$ to replace $\mathrm{Tr}\{{\mathbf{C}}_{\mathrm{e}}[n]\}$ in the objective function in \eqref{Eqn:ProblemFormulation}, the problem can be equivalently expressed as follows
\begin{align}\label{Eqn:ProblemFormulationRA}
\underset{{u}_k [n], \mathbf{w}_k [n], \mathbf{w}_{\mathrm{e}} [n],\xi[n]}{\mathrm{maximize}}\,\; & \frac{1}{1+\omega}\frac{\sum\nolimits_{k=1}^{K} u_k[n]}{K} - \frac{\omega}{1+\omega} \frac{\xi[n]}{\mathrm{MSE}_{\mathrm{max}}}
\\[-1mm]
\notag\mbox{s.t.}\;\;
&\mbox{{C1-C3}}, \notag\\[-2mm]
&{\mbox{C4}}:\; \mathrm{Tr}\{{\mathbf{C}}_{\mathrm{e}}[n]\} \le \xi[n].\notag
\end{align}
\vspace{-10mm}\par\noindent
Further introducing auxiliary variables $t_i \ge 0$, $\forall i \in \left\{1,\ldots,6\right\}$, to bound the diagonal entries of the posterior covariance matrix ${\mathbf{C}}_{\mathrm{e}}[n]$, i.e., $\{{\mathbf{C}}_{\mathrm{e}}[n]\}_{ii} \le t_i$, constraint ${\mbox{C4}}$ is equivalently rewritten as
\vspace{-1mm}
\begin{equation}
	{\mbox{C4a}}:\;\sum_{i=1}^{6}t_i \le \xi[n]\;\;\text{and}\;\;{\mbox{C4b}}:\;\left[ {\begin{array}{*{20}{c}}
		{{\bf{C}}_{\rm{e}}^{ - 1}[n]}&{{{\bf{e}}_i}}\\
		{{\bf{e}}_i^{\rm{T}}}&{{t_i}}
		\end{array}} \right] \succeq {\bf{0}}, \forall i,\vspace{-1mm}
\end{equation}
where ${{{\bf{e}}_i}} \in \mathbb{R}^{6 \times 1}$ is the $i$-th column of $\mathbf{I}_6$.
Moreover, due to the channel prediction uncertainty, there are infinitely many possibilities for constraint ${\mbox{C3}}$.
Adopting the $\mathcal{S}$-Procedure \cite{ZhiqiangSWIPT}, constraint ${\mbox{C3}}$ can be restricted to the following linear matrix inequality (LMI):
\vspace{-1mm}
\begin{equation}\label{Eqn:C3LMI}
\overline{\mbox{C3}}:\; u_k[n]\left[ {\begin{array}{*{20}{c}}
	{{\epsilon _{k}}\mathbf{I}_{M^{\mathrm{t}}_{\mathrm{b}}} -\mathbf{a}_{2,k}[n]}&-\mathbf{a}^{\mathrm{H}}_{2,k}[n] \hat{{\bf{h}}}_{{\rm{be}}}[n|n-1]\\
	-\hat{{\bf{h}}}^{\mathrm{H}}_{{\rm{be}}}[n|n-1]\mathbf{a}_{2,k}[n] &{ -{\epsilon _{k}}\sigma^2_{\hat{\mathbf{h}}_{\mathrm{be}}[n|n\hspace{-0.25mm}-\hspace{-0.25mm}1]} - c_{2,k}[n]}
	\end{array}} \right] \succeq {\bf{0}},\vspace{-1mm}
\end{equation}
where ${\epsilon _{k}} >0$ is an auxiliary variable, $\mathbf{a}_{2,k}[n] = {{\bf{W}}_k[n]} - \left(2^{R_{\mathrm{Leakage}}}-1\right) {{\bf{W}}_{\mathrm{e}}[n]}$, $c_{2,k}[n] = \hat{{\bf{h}}}^{\mathrm{H}}_{{\rm{be}}}[n|n-1] \mathbf{a}_{2,k}[n] \hat{{\bf{h}}}_{{\rm{be}}}[n|n-1] - \left(2^{R_{\mathrm{Leakage}}}-1\right)\frac{\sigma _{\rm{e}}^2}{M_{\mathrm{e}}}$, ${{{\bf{W}}_k}[n]} = {{{\bf{w}}_k}[n]}{{{\bf{w}}^{\mathrm{H}}_k}[n]}$, and ${{{\bf{W}}_{\mathrm{e}}}[n]} = {{{\bf{w}}_{\mathrm{e}}}[n]}{{{\bf{w}}^{\mathrm{H}}_{\mathrm{e}}}[n]}$.
Now, defining ${{\bf{H}}_{k}^{}[n]} = {{\bf{h}}_{k}^{}[n]}{{\bf{h}}_{k}^{\mathrm{H}}[n]}$, the problem in \eqref{Eqn:ProblemFormulationRA} can be rewritten as follows
\vspace{-1mm}
\begin{align}\label{Eqn:ProblemFormulationRAII}
&\underset{{u}_k [n],\mathbf{w}_k [n],\mathbf{w}_{\mathrm{e}} [n],t_i,{\epsilon _{k}},\xi[n]}{\maxo}\,\, \frac{1}{1+\omega}\frac{\sum\nolimits_{k=1}^{K} u_k[n]}{K} - \frac{\omega}{1+\omega} \frac{\xi[n]}{\mathrm{MSE}_{\mathrm{max}}}\\[-1mm]
\notag\mbox{s.t.}\;\;
{\mbox{{C1}}}:\; &\sum\nolimits_{k = 1}^{K} u_k[n]\Tr\left(\mathbf{W}_{k}\left[ n \right]\right) + \Tr\left(\mathbf{W}_{\mathrm{e}}\left[ n \right]\right) \le p_{\mathrm{max}},\notag\\[-1mm]
{\mbox{{C2}}}:\; & u_k[n]{\rm{Tr}}\left( {{{\bf{W}}_k}[n]{\bf{H}}_{k}^{}[n]} \right) \notag\\[-1mm]
&\ge \left( {{2^{{u_k[n]R_{{\rm{min}}}}}} - 1} \right)\left[ \sum\nolimits_{k' \ne k}u_{k'}[n] {{\rm{Tr}}\left( {{{\bf{W}}_{k'}}[n]{\bf{H}}_{k}^{}[n]} \right)}  + {\rm{Tr}}\left( {{{\bf{W}}_{\rm{e}}}[n]{\bf{H}}_{k}^{}[n]} \right) + {\sigma _k^2} \right], \forall k,\notag\\[-2mm]
\overline{\mbox{C3}},\; \;&{\mbox{{C4a}}},{\mbox{{C4b}}},\notag\\[-2mm]
{\mbox{{C5}}}:\; & \Rank\left({{{\bf{W}}_k}[n]}\right) \le 1, \forall k, \Rank\left({{{\bf{W}}_{\mathrm{e}}}[n]}\right) \le 1.\notag
\end{align}
\vspace{-10mm}\par\noindent
The main obstacles for solving \eqref{Eqn:ProblemFormulationRAII} are the binary variables ${u}_k [n]$ and the non-convex rank-one constraint C5.
Besides, we can observe that given the user scheduling variables, the problem in \eqref{Eqn:ProblemFormulationRAII} is a convex optimization problem except for constraint C5, which
can be addressed by the commonly-used SDR approach \cite{ZhiQuanTSPM,ZhiqiangSWIPT}.
However, finding the optimal user scheduling strategy requires an exhaustive search over all $2^K$ user scheduling possibilities, which has a prohibitively high computational complexity.
Thus, to facilitate online resource allocation design, we propose a low-complexity channel correlation-based user scheduling algorithm and an SDR-based precoding design.

\subsubsection{SDR-based Precoding Design}
Given the user scheduling variables, the problem in \eqref{Eqn:ProblemFormulationRAII} is a tracking MSE minimization problem.
After removing rank-one constraint C5, the resulting problem is a convex semidefinite programming (SDP) problem:
\vspace{-2mm}
\begin{equation}\label{Eqn:ProblemFormulationRAIII}
\underset{\mathbf{w}_k [n],\mathbf{w}_{\mathrm{e}} [n],t_i,{\epsilon _{k}},\xi[n]}{\mino}\,\,  \xi[n]\;\;\mbox{s.t.}\;\;
{\mbox{{C1}}},{\mbox{{C2}}},\overline{\mbox{C3}}, {\mbox{{C4a}}},{\mbox{{C4b}}},\vspace{-1mm}
\end{equation}
which can be solved by numerical convex solvers, such as CVX\cite{book:convex}.
We can show that the solution of \eqref{Eqn:ProblemFormulationRAIII} satisfies constraint C5 in \eqref{Eqn:ProblemFormulationRAIII}, i.e., the SDR is tight.
The corresponding proof closely follows a similar proof in the appendix of \cite{ZhiqiangSWIPT} and is omitted here.
The precoding vectors $\mathbf{w}_k [n]$ and $\mathbf{w}_{\mathrm{e}} [n]$ can be obtained from the eigen-vectors of $\mathbf{W}_k [n]$ and $\mathbf{W}_{\mathrm{e}} [n]$ associated with the maximum eigen-values, respectively.

\subsubsection{Channel Correlation-based User Scheduling}
The channel correlation coefficient between the communication channel of GU $k$ and the wiretap channel of the E-UAV is defined as 
\vspace{-1mm}
\begin{equation}
\rho^{k}_{\mathrm{e}} = \frac{\left|{{\bf{h}}_{k}^{\mathrm{H}}[n]}{\hat{\mathbf{h}}_{\mathrm{be}}[n|n\hspace{-0.25mm}-\hspace{-0.25mm}1]}\right|}{\left\|{{\bf{h}}_{k}^{\mathrm{H}}[n]}\right\|\left\|{\hat{\mathbf{h}}_{\mathrm{be}}[n|n\hspace{-0.25mm}-\hspace{-0.25mm}1]}\right\|}.\vspace{-1mm}
\end{equation}
In general, the higher channel correlation coefficients $\rho^{k}_{\mathrm{e}}$ is, the higher the risk of leaking the information of GU $k$. 
Therefore, selecting the users with lower channel correlation coefficients for service is beneficial for secure communications and helps save power to improve the sensing performance.  
Without loss of generality, the GUs are indexed in ascending order of their channel correlation coefficients, i.e., $\rho^{1}_{\mathrm{e}}\le \rho^{2}_{\mathrm{e}} \le \ldots \le \rho^{K}_{\mathrm{e}}$.
Then, we schedule the first $\tilde{K}$ users for service, i.e., $u_k[n] = 1$, $\forall k \le \tilde{K}$, and solve the problem in \eqref{Eqn:ProblemFormulationRAIII} $(K+1)$ times to obtain the optimal objective values, i.e., $\mathrm{Obj}(\tilde{K})$, $\tilde{K} = 0,\ldots,K$.
The proposed user scheduling strategy is as follows:
\vspace{-1mm}
\begin{equation}
	{u_k}[n] = \left\{ \begin{array}{l}
	1,\;k \le \tilde{K}^*,\\[-1mm]
	0,\;k > \tilde{K}^*,
	\end{array} \right.\vspace{-1mm}
\end{equation}
where $\tilde{K}^* =\arg\max \mathrm{Obj}(\tilde{K})$.
Note that there are only $K+1$ possible user scheduling strategies to compare and $\mathrm{Obj}(\tilde{K})$, $\tilde{K} = 0,\ldots,K$, can be obtained in parallel.

\vspace{-4mm}
\section{Simulation Results}
\label{Sec:SimlationResults}
In this section, we use computer simulations to evaluate the performance and to provide insights regarding the operation of the proposed ISNC framework.
Unless specified otherwise, we adopt the system parameters given in Table \ref{simulation_setting} in our simulations.
The initial state of the I-UAV is given by ${\bf{q}}_{\mathrm{b}}\left[ 1 \right]$, $\dot{\bf{q}}_{\mathrm{b}}\left[ 1 \right]$, $\ddot{\bf{q}}_{\mathrm{b}}\left[ 1 \right]$ and the locations of the $K$ GUs are shown in Fig. \ref{fig:NavigationandSensingSingleStep}.
The E-UAV cruises over all GUs with the minimum route length and at the minimum allowed altitude, see Fig. \ref{fig:NavigationandSensingSingleStep}.
The corresponding trajectory design is an instance of the well-known traveling salesman problem and is solved by a genetic algorithm\cite{GAAlgorithm}.
\begin{table*}[t]
	\scriptsize
	\vspace{-5mm}
	\caption{Simulation parameters.} \label{simulation_setting}
	\vspace{-7mm}
	\begin{center}
		\begin{tabular}{ c | c |c |c | c | c | c| c}
			\hline			
			Parameter         & Value & Parameter          & Value      & Parameter         & Value  &   Parameter          & Value \\ \hline
			$K$               &      $10$           &   $[M^{\mathrm{tx}}_{\rm{b}},M^{\mathrm{tx}}_{\rm{b}}]$  &       $[4,4]$       &    $[M^{\mathrm{rx}}_{\rm{b}},M^{\mathrm{rx}}_{\rm{b}}]$  &       $[4,4]$    & $[M^{\mathrm{x}}_{\rm{e}},M^{\mathrm{x}}_{\rm{e}}]$  &       $[4,4]$    \\
			$\delta$        & $1$ s              &   $N$               &       187       &$V_{\mathrm{max}}$ & $30$ m/s            &   $[A^x_{\mathrm{cc}},A^y_{\mathrm{cc}},A^z_{\mathrm{cc}}]$  &     $[4, 4, 2] \;\mathrm{m/s^2}$    \\
			$\beta_0$         &  $-50$ dBW & $\mathbf{q}_{\mathrm{min}}$ & $[0,0,50]^{\mathrm{T}}$ & $\mathbf{q}_{\mathrm{max}}$ & $[500,500,100]^{\mathrm{T}}$ & $\vartheta_{\mathrm{e}}$&  $0.1\; \mathrm{m^2}$ \\
			$R_{\mathrm{min}}$  &    $5$ bit/s/Hz       &   $R_{\mathrm{Leakage}}$  &    $0.01$ bit/s/Hz             &     $\mathrm{MSE}_{\mathrm{max}} $       &   $10$               &      $d_{\mathrm{min}}$&  $10$ m  \\
			$\sigma^2_k = \sigma^2_{\mathrm{e}}$  &    $-99$ dBm  & $\sigma^2_{\mathrm{b}}$  &    $-50$ dBm     &   $f_{\mathrm{c}}$  &  $3$ GHz               & $c$               &   $3 \times 10^8$ m/s      \\  $p_{\mathrm{max}}$ &  $30$ dBm 
			  &   $G_{\mathrm{MF}}$           &   $10^4$ & $c_{{\tau _{\rm{e}}}}$     &  $10^{-6}$              &  $c_{\nu _{\rm{e}}}$     &  $50$  \\
			$c_{\theta_{\mathrm{be}}}$ & $0.1$    &    $c_{\phi_{\mathrm{be}}}$      &  $0.1$    & $ \lambda$              &   $\left(-1,1\right)$              &  $\omega$ & $\left[0, \infty \right)$ \\
			 ${\bf{q}}_{\mathrm{b}}\left[ 1 \right]$ & $\frac{\mathbf{q}_{\mathrm{min}}+\mathbf{q}_{\mathrm{max}}}{2}$ & $\dot{\bf{q}}_{\mathrm{b}}\left[ 1 \right]$ & $[0,0,0]^{T}$  & $\ddot{\bf{q}}_{\mathrm{b}}\left[ 1 \right]$  &$[0,0,0]^{T}$ & $\left[\delta_{\mathrm{Sen}},\;\delta_{\mathrm{Com}}\right]$& $\left[0.3,\;0.7\right]$\\\hline
		\end{tabular}
	\end{center}
\vspace{-10mm}
\end{table*}

\vspace{-4mm}
\subsection{Navigation and Sensing Performance}

\begin{figure}[t]
	\begin{minipage}{.49\textwidth}
		\center{\includegraphics[width=3.5in]{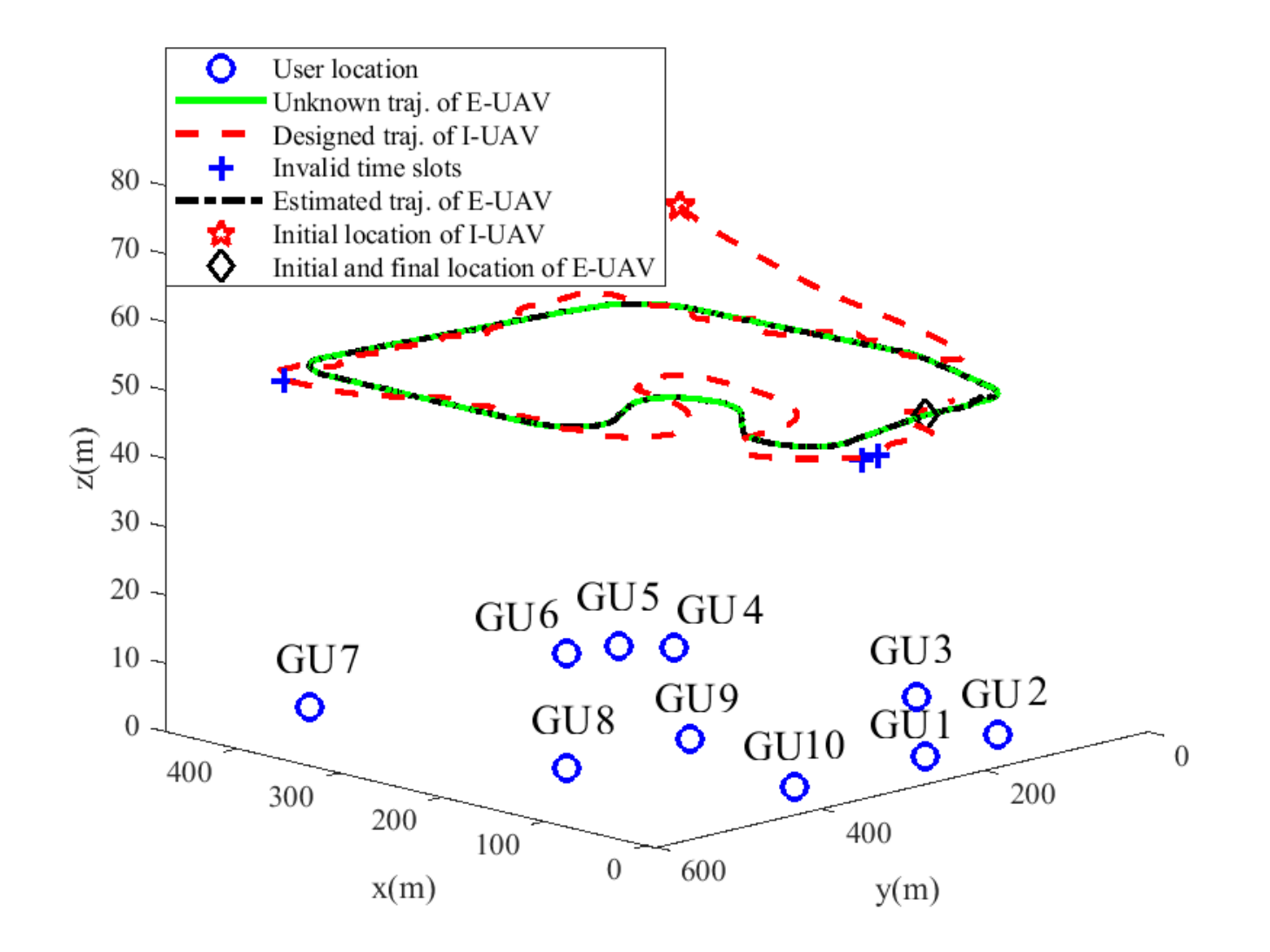}}\vspace{-10mm}
		\caption{Designed trajectory of I-UAV and estimated trajectory of E-UAV for the proposed scheme with $L=1$.}\vspace{-10mm}
		\label{fig:NavigationandSensingSingleStep}
	\end{minipage}
	\hspace{1mm}
	\begin{minipage}{.49\textwidth}
		\center{\includegraphics[width=3.5in]{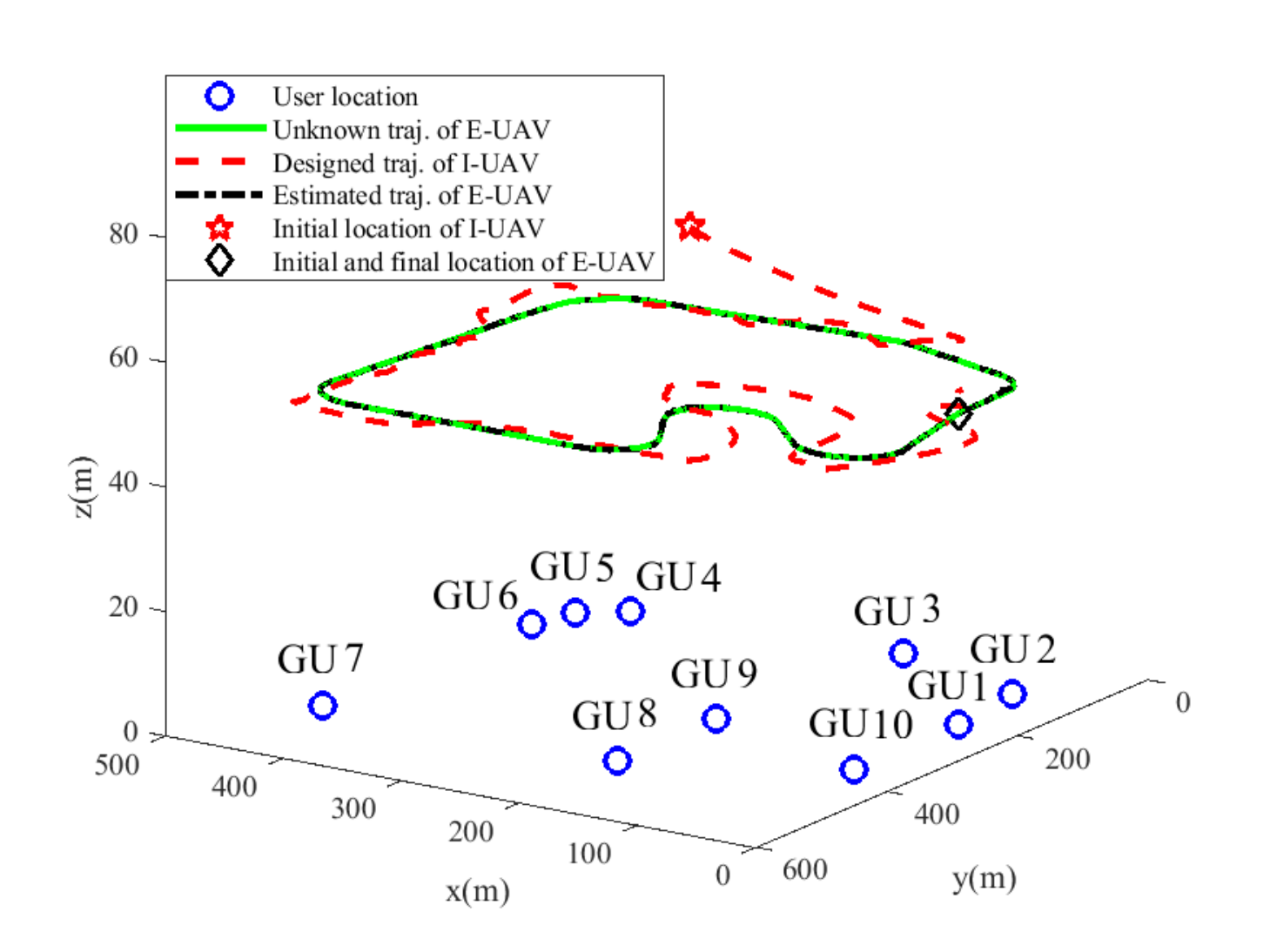}}\vspace{-10mm}
		\caption{Designed trajectory of I-UAV and estimated trajectory of E-UAV for the proposed scheme with $L=5$.}\vspace{-10mm}
		\label{fig:NavigationandSensingMultiStep}	
	\end{minipage}
\end{figure}

The designed trajectory of the I-UAV and the estimated trajectory of the E-UAV for the proposed ISNC scheme are shown in Figs. \ref{fig:NavigationandSensingSingleStep} and \ref{fig:NavigationandSensingMultiStep} for the proposed single-step ($L=1$) and multi-step ($L=5$) navigation designs, respectively.
The navigation parameter to determine the desired destination point is set as $\lambda = 0.95$.
As can be observed, the estimated trajectory aligns perfectly with the unknown trajectory of the E-UAV for both $L=1$ and $L=5$.
This is because the navigation policy of the I-UAV is designed to approach the E-UAV for $\lambda \to 1$, which is beneficial for sensing.
Besides, there are three invalid time slots in Fig. \ref{fig:NavigationandSensingSingleStep} for the proposed single-step navigation design, i.e., when $L=1$, the navigation design problem in \eqref{Eqn:NavProblemFormulationII} is infeasible for $n = [99, 169, 170]$, which is caused by the inherent momentum of the I-UAV.
In contrast, the proposed multi-step navigation design in Fig. \ref{fig:NavigationandSensingMultiStep} is always valid as the I-UAV can predictively identify whether it is approaching the boundary or the E-UAV and alter its direction to avoid violating constraints C3 and C4 in \eqref{Eqn:NavProblemFormulationII} in the subsequent $L$ time slots.

\begin{figure}[t]
	\vspace{-5mm}
	\begin{minipage}{.49\textwidth}
		\center{\includegraphics[width=3.5in]{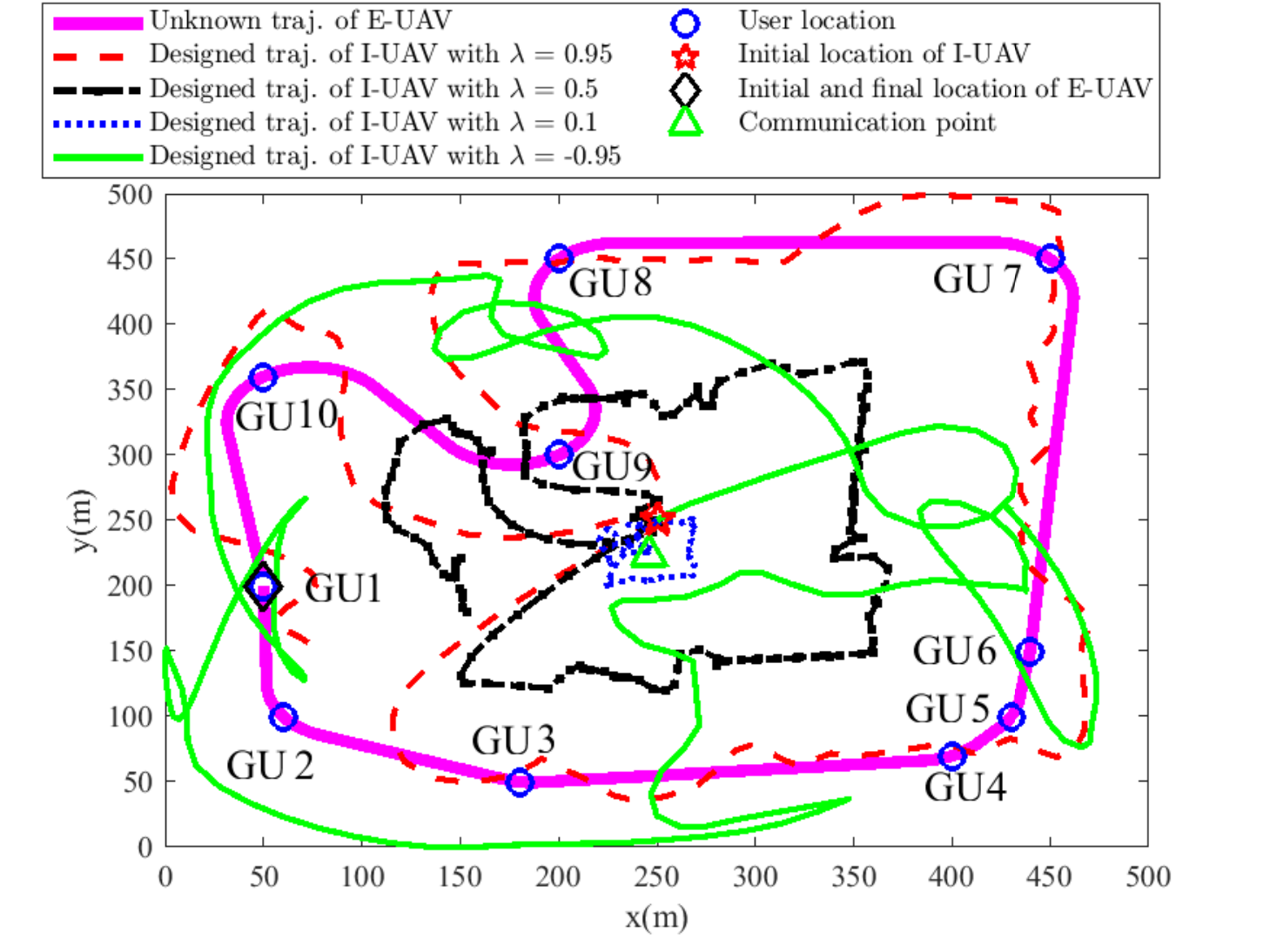}}\vspace{-7mm}
		\caption{Designed trajectory of I-UAV for the proposed ISNC scheme for different $\lambda$.}\vspace{-7mm}
		\label{fig:NavigationMultiStep_DifferentLambda}	
	\end{minipage}
	\hspace{1mm}
	\begin{minipage}{.49\textwidth}
		\center{\includegraphics[width=3.5in]{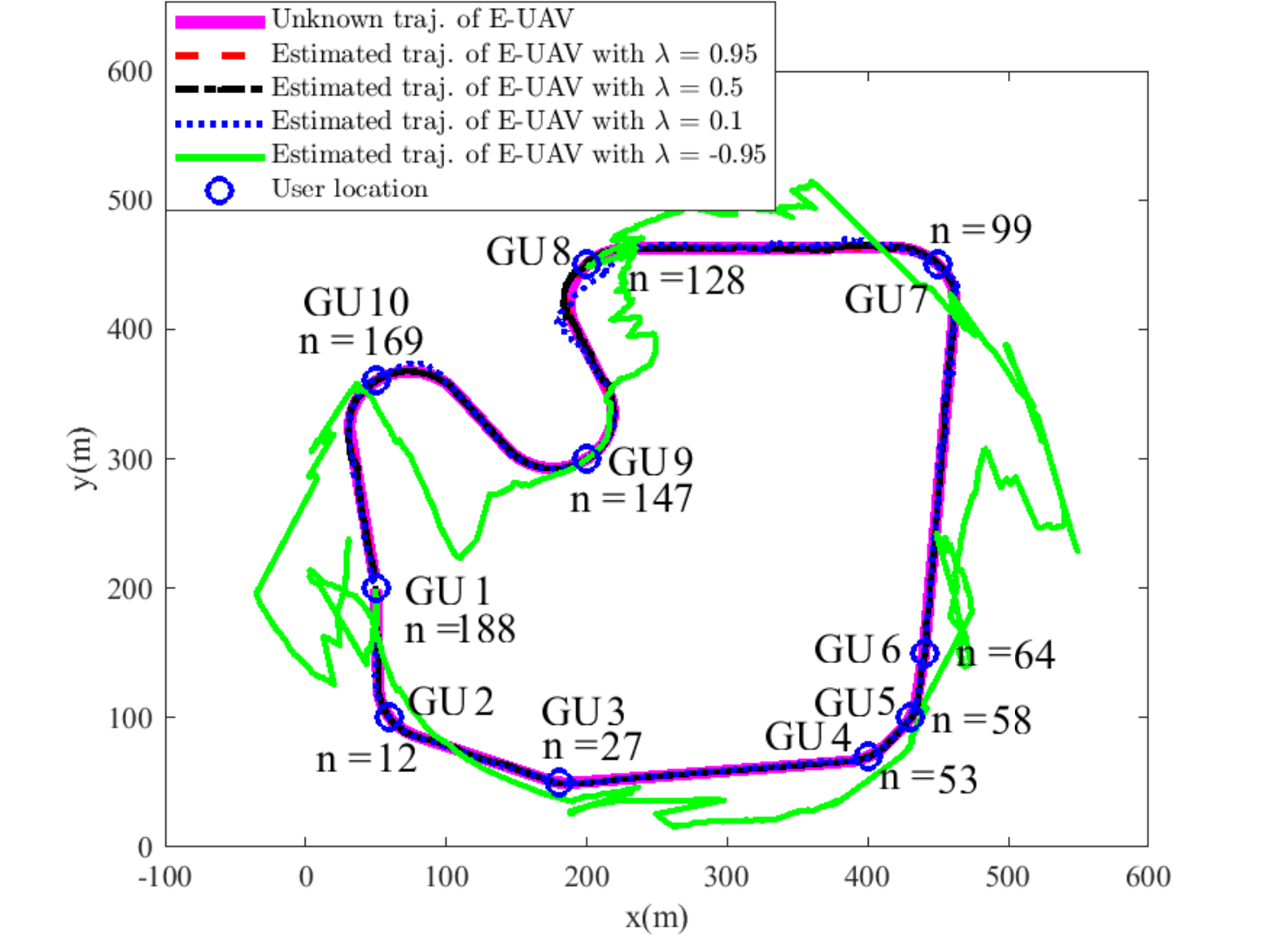}}\vspace{-7mm}
		\caption{Estimated trajectory of E-UAV for the proposed ISNC scheme for different $\lambda$.}\vspace{-7mm}
		\label{fig:SensingMultiStep_DifferentLambda}	
	\end{minipage}
\end{figure}

\begin{figure}[t]
	\begin{minipage}{.49\textwidth}
		\center{\includegraphics[width=3.5in]{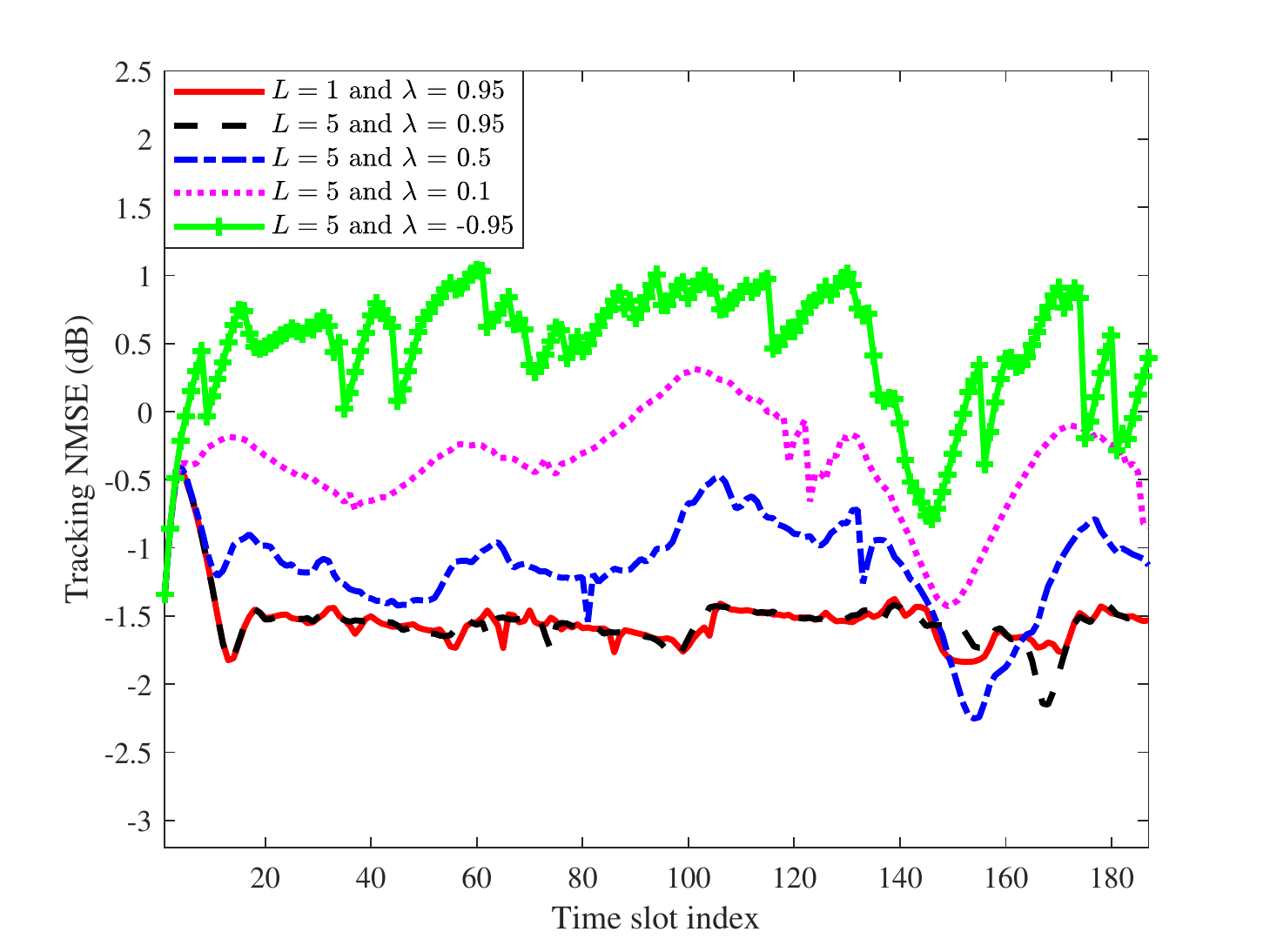}}\vspace{-7mm}
		\caption{Tracking NMSE of E-UAV state for the proposed ISNC scheme for different $L$ and $\lambda$.}\vspace{-10mm}
		\label{fig:TrackingMSEMultiStep_DifferentLambda}	
	\end{minipage}
	\hspace{1mm}
	\begin{minipage}{.49\textwidth}
		\center{\includegraphics[width=3.5in]{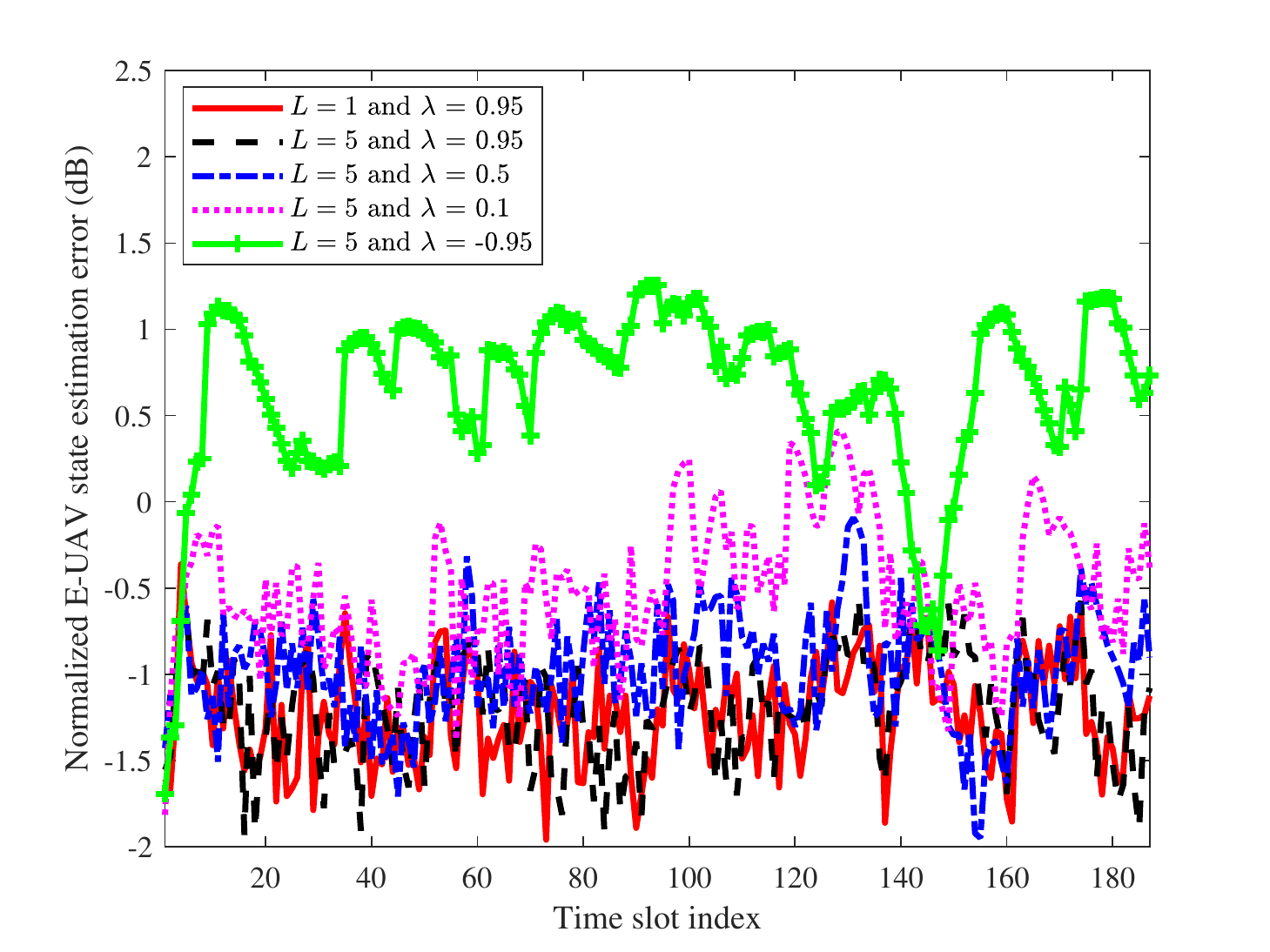}}\vspace{-7mm}
		\caption{E-UAV state estimation error for the proposed ISNC scheme for different $L$ and $\lambda$.}\vspace{-10mm}
		\label{fig:EstimatedErrorMultiStep_DifferentLambda}	
	\end{minipage}
\end{figure}

The designed trajectory of the I-UAV and the estimated trajectory of the E-UAV for the proposed ISNC scheme and different navigation parameters $\lambda$ are depicted in Figs. \ref{fig:NavigationMultiStep_DifferentLambda}
and \ref{fig:SensingMultiStep_DifferentLambda}, respectively.
The corresponding normalized posterior MSE (NMSE) for E-UAV state tracking $\log_{10}\left(\frac{\Tr\{{\mathbf{C}}_{\mathrm{e}}[n]\}}{\mathrm{MSE}_{\mathrm{max}}}\right)$ and the instantaneous normalized E-UAV state estimation error $\log_{10}\left(\frac{\left|{{\boldsymbol{\alpha}}}_{\mathrm{e}} [n] - \hat{{\boldsymbol{\alpha}}}_{\mathrm{e}} [n]\right|^2}{\mathrm{MSE}_{\mathrm{max}}}\right)$ are shown in Figs. \ref{fig:TrackingMSEMultiStep_DifferentLambda}
and \ref{fig:EstimatedErrorMultiStep_DifferentLambda}, respectively.
We can observe in Fig. \ref{fig:NavigationMultiStep_DifferentLambda} that when $\lambda$ is decreased from $0.95$ to $0.1$, the designed trajectory of the I-UAV alters from following the E-UAV to staying close to the communication point.
For $\lambda = -0.95$, the designed trajectory of the I-UAV is opposite to that of the E-UAV since navigation parameters $\lambda\to -1$ cause the I-UAV to fly away from the E-UAV.
Yet, decreasing $\lambda$ degrades the sensing performance, as evident from Figs. \ref{fig:SensingMultiStep_DifferentLambda}, \ref{fig:TrackingMSEMultiStep_DifferentLambda}, and  \ref{fig:EstimatedErrorMultiStep_DifferentLambda}, since the I-UAV is farther from the E-UAV and the backscattered AN suffers a significant round-trip path loss.
Besides, in Fig. \ref{fig:SensingMultiStep_DifferentLambda}, we can observe that the estimated trajectory of the E-UAV deviates from the ground-truth for $\lambda = [0.1, -0.95]$, especially when the E-UAV is at the turning points in time slots $n=[99, 128, 169]$.
In fact, decreasing $\lambda$ leads to a lower MF output SNR in \eqref{Eqn:MFsnr} and thus the uncertainty of the measurements in \eqref{Eqn:MeautrementModel_II} increases.
In this power-limited regime, the EKF is more sensitive to the state evolution model mismatch of \eqref{Eqn:StateEvolution}, when the I-UAV is turning.
Moreover, in Figs. \ref{fig:TrackingMSEMultiStep_DifferentLambda} and \ref{fig:EstimatedErrorMultiStep_DifferentLambda}, it can be seen that for $\lambda = 0.95$, the tracking NMSE of the E-UAV state and the instantaneous normalized E-UAV state estimation error are both small for both the single-step and multi-step navigation designs.
In fact, navigation parameters $\lambda\to 1$ cause the I-UAV to closely chase the E-UAV and thus the designed trajectory of the I-UAV aligns well with that of the E-UAV, as illustrated in Figs. \ref{fig:NavigationandSensingSingleStep} and \ref{fig:NavigationandSensingMultiStep}.
Therefore, for $\lambda = 0.95$, both the single-step and multi-step navigation designs can efficiently exploit the backscattered AN for sensing and result in a superior sensing performance.
%
%
Comparing Figs. \ref{fig:TrackingMSEMultiStep_DifferentLambda} and \ref{fig:EstimatedErrorMultiStep_DifferentLambda}, it can be observed that the tracking NMSE of the E-UAV state and the instantaneous normalized E-UAV state estimation error follow a similar trend but are not aligned with each other exactly.
This is because E-UAV state tracking suffers not only from measurement uncertainty but also from state evolution model mismatch. 
For instance, for $\lambda = 0.1$, around time slot $n = 128$, the I-UAV is close to the E-UAV but the E-UAV is turning.
Therefore, the tracking NMSE of the E-UAV state in Fig. \ref{fig:TrackingMSEMultiStep_DifferentLambda} is small, while the instantaneous the E-UAV state estimation error in Fig. \ref{fig:EstimatedErrorMultiStep_DifferentLambda} is large.

\vspace{-4mm}
\subsection{Secure Communication Performance}

\begin{figure}[t]
	\vspace{-5mm}
	\begin{minipage}{.49\textwidth}
		\center{\includegraphics[width=3.5in]{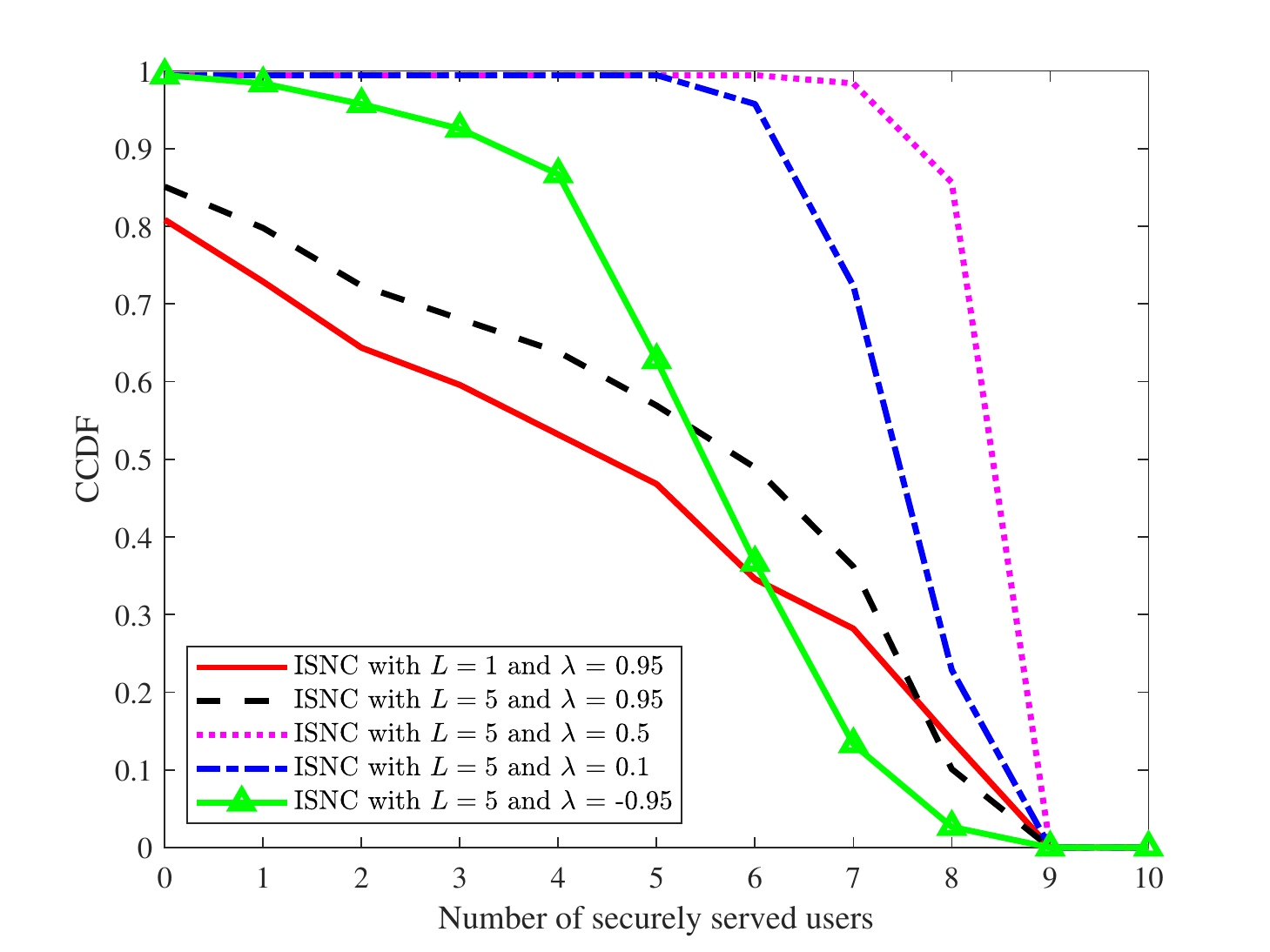}}\vspace{-7mm}
		\caption{The CCDF of the number of users securely served for the proposed ISNC scheme for different $L$ and $\lambda$.}\vspace{-10mm}
		\label{fig:ComPerformanceMultiStep_DifferentLambda}	
	\end{minipage}
	\hspace{1mm}
	\begin{minipage}{.49\textwidth}
		\center{\includegraphics[width=3.5in]{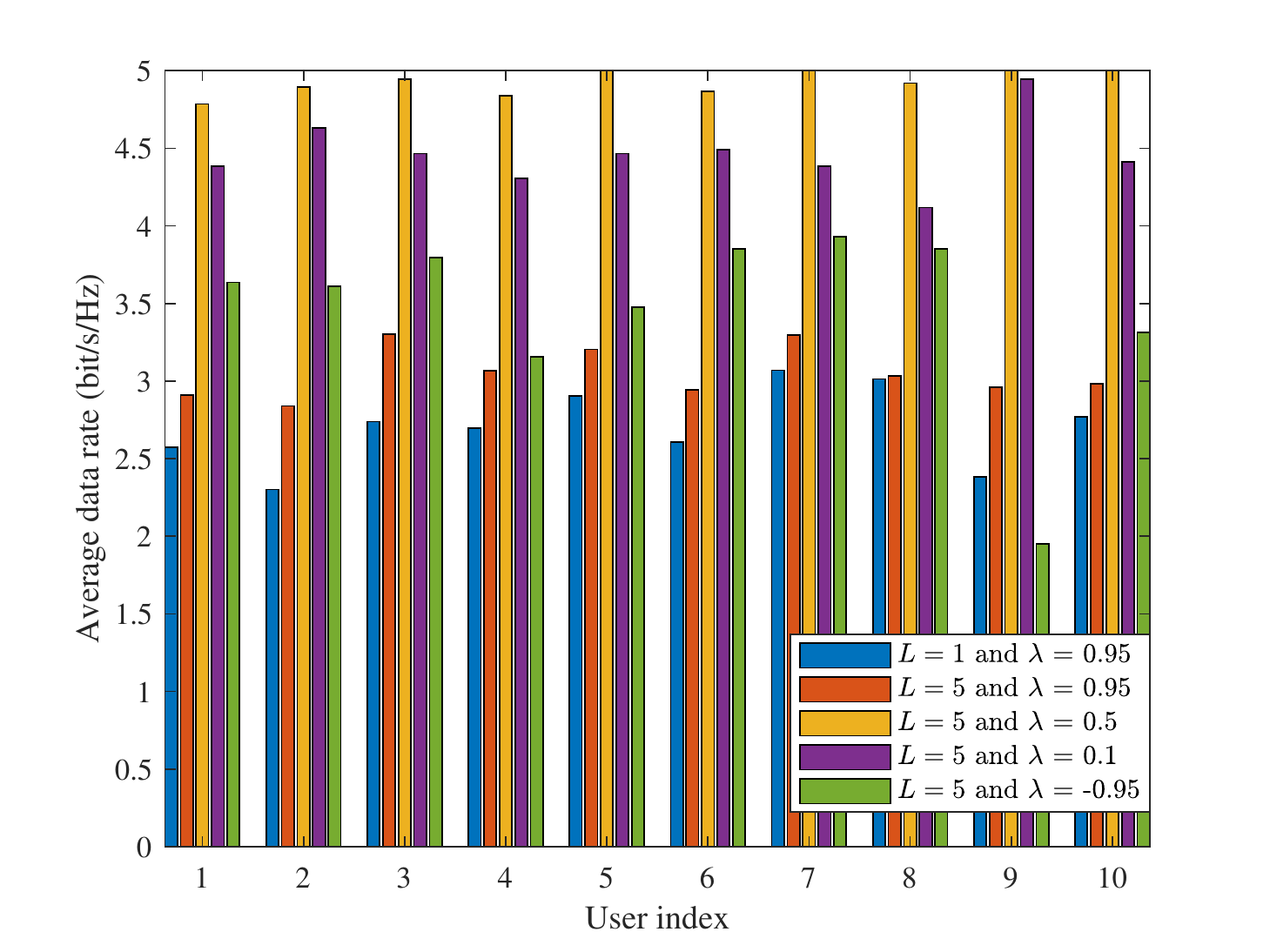}}\vspace{-7mm}
		\caption{Average data rate of each user for the proposed ISNC scheme for different $L$ and $\lambda$.}\vspace{-10mm}
		\label{fig:RatePerformanceMultiStep_DifferentLambda}	
	\end{minipage}
\end{figure}

In this section, we evaluate the secure communication performance of the proposed ISNC scheme.
In Fig. \ref{fig:ComPerformanceMultiStep_DifferentLambda}, for different $L$ and $\lambda$, we depict the empirical complementary cumulative distribution function (CCDF) of the number of users securely served during $N$ time slots, i.e., $F(m) = \sum_{n=1}^{N}I\{\sum\nolimits_{k=1}^{K} u_k[n] > m\}/N$, where $I\{\cdot\}$ is the indicator function.
Note that the empirical CCDF decreases with increasing $m$ and $F(0)<1$ implies that no user can be served securely in some of the $N$ time slots.
Also, the higher the curve in Fig. \ref{fig:ComPerformanceMultiStep_DifferentLambda}, the better the secure communication performance.
In Fig. \ref{fig:RatePerformanceMultiStep_DifferentLambda}, we show the average data rate of all $K$ users for the proposed scheme for different $L$ and $\lambda$.
In Fig. \ref{fig:ComPerformanceMultiStep_DifferentLambda}, we can observe that only part of the GUs can be securely served in all time slots in all considered cases, as $F(9) = 0$ for all curves.
As a result, the average data rate of each user is strictly smaller than the minimum required instantaneous data rate ${R_{\mathrm{min}}} = 5$ bit/s/Hz in Fig. \ref{fig:RatePerformanceMultiStep_DifferentLambda}.
Also, it can be seen that the secure communication performances for both $\lambda = 0.95$ and $\lambda = -0.95$ are worse than that for $\lambda = \{0.5, 0.1\}$.
In fact, having the I-UAV closely approach the E-UAV, as is the case for $\lambda \to 1$, results in the wiretap channel being stronger than the downlink communication channels to the legitimate GUs, which are far away from the I-UAV, leading to a poor secure communication performance.
On the other hand, having the I-UAV fly away from the E-UAV, as is the case for $\lambda \to -1$, results in a degraded sensing performance as mentioned before and the quality of the wiretap channel prediction in \eqref{Eqn:PredictedChannel} decreases.
As a result, the resource allocation design in \eqref{Eqn:ProblemFormulation} for broadcasting information to the GUs becomes more conservative and more power has to be allocated to AN for jamming and sensing.
Moreover, it can be seen that the secure communication performance of the proposed ISNC scheme is higher for $\lambda = 0.5$ than for $\lambda = 0.1$.
This is because a small $\left|\lambda\right|$ does not exploit the mobility of the UAV for improving communication performance, see Fig. \ref{fig:NavigationMultiStep_DifferentLambda}.
Additional discussion on the trade-off between sensing and communication performance will be provided in Section IV-D.

\vspace{-4mm}
\subsection{Comparison with Baseline Schemes}
\begin{figure}[t]
	\vspace{-5mm}
	\center{\includegraphics[width=3.8in]{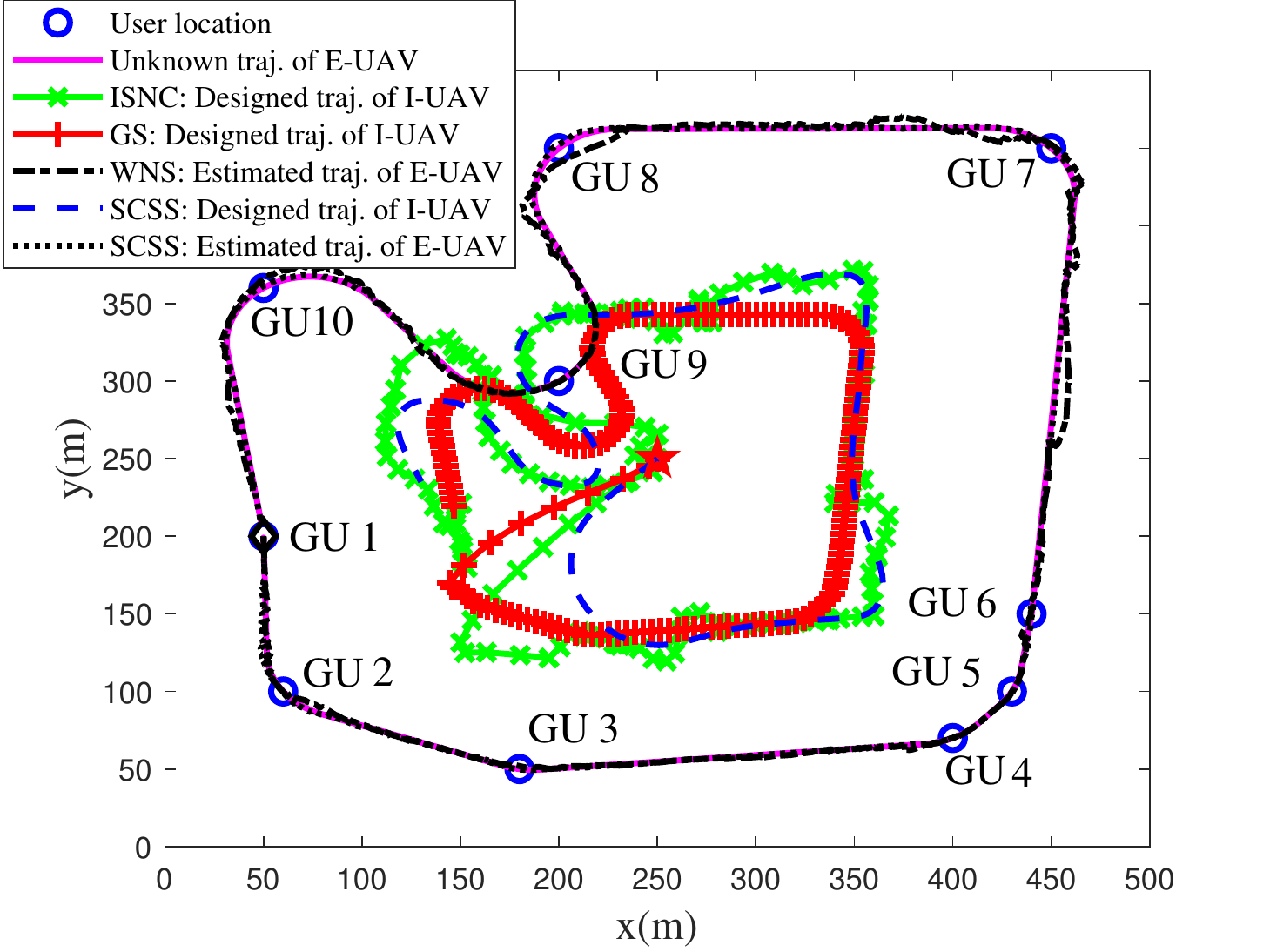}}\vspace{-7mm}
	\caption{Designed trajectory of I-UAV and estimated trajectory of E-UAV for baseline schemes.}\vspace{-10mm}
	\label{fig:NavigationandSensingBaseline}
\end{figure}

In this section, three baseline schemes are considered to demonstrate the benefits of integrating sensing and navigation functionalities for secure communications in UAV wireless networks: 1) \textbf{Genius scheme (GS)}, where the I-UAV knows the E-UAV state perfectly and thus the wiretap channel prediction in \eqref{Eqn:PredictedChannel} is also perfect, i.e., $\sigma^2_{\hat{\mathbf{h}}_{\mathrm{be}}[n|n\hspace{-0.25mm}-\hspace{-0.25mm}1]} = 0$; 2) \textbf{Without navigation scheme (WNS)}, in which the I-UAV is fixed at its initial location but can sense the state of the E-UAV and predict the wiretap channel for resource allocation design; 3) \textbf{Separated communication and sensing scheme (SCSS)}, in which we perform communications and sensing at the I-UAV in a time sharing manner.
For ``SCSS'', the highlighted part in Fig. \ref{fig:SlotStructure} is divided into two parts, i.e., one for flight and communication and the other one for flight and sensing.
A fraction of time $\delta_{\mathrm{Com}}$ is used for secure communication and the remaining time $\delta_{\mathrm{Sen}}$ is used for sensing, where $\delta_{\mathrm{Com}} + \delta_{\mathrm{Sen}} = \delta$.
The designs proposed in this paper are applicable for separated communication and sensing scheme with some modifications.
For communication, the achievable data rate in \eqref{Eqn:DownlinkRate} is reduced by a ratio of $\frac{\delta_{\mathrm{Com}}}{\delta}$.
Then, the resource allocation design in \eqref{Eqn:ProblemFormulation} does not need to consider sensing performance, i.e., $\omega = 0$.
For sensing, the MF output SNR in \eqref{Eqn:MFsnr} has to be modified as
\begin{equation}\label{Eqn:MFsnrII}
	\mathrm{SNR} \hspace{-1mm}=\hspace{-1mm} \frac{\vartheta_{\mathrm{e}}\beta^2_0\frac{\delta_{\mathrm{Sen}}}{\delta}G_{\mathrm{MF}}M^{\mathrm{r}}_{\mathrm{b}} \big|\mathbf{a}^{\mathrm{H}}_{{M^{\mathrm{tx}}_{\mathrm{b}}},{M^{\mathrm{ty}}_{\mathrm{b}}}}\left(\theta_{\mathrm{e}}[n], \phi_{\mathrm{e}}[n]\right) \mathbf{w}_{\mathrm{Sen}} [n]\big|^2}{16\pi \sigma_{\mathrm{b}}^2d^4_{\mathrm{e}}[n]},
\end{equation}
where the MF gain $G_{\mathrm{MF}}$ is reduced by a factor $\frac{\delta_{\mathrm{Sen}}}{\delta}$.
The dedicated precoding vector for sensing $\mathbf{w}_{\mathrm{Sen}} [n]$ is obtained based on the predicted state of the E-UAV and is given by
\vspace{-1mm}
\begin{equation}\label{eqn:SensingVectorSCSS}
	\mathbf{w}_{\mathrm{Sen}} [n] = \sqrt{{p_{\mathrm{max}}}/{M_{\mathrm b}^{\mathrm{t}}}} \mathbf{a}_{{M^{\mathrm{tx}}_{\mathrm{b}}},{M^{\mathrm{ty}}_{\mathrm{b}}}}\hspace{-1mm}\left(\hspace{-0.5mm}\hat{\theta}_{\mathrm{e}}\hspace{-0.25mm}[\hspace{-0.25mm}n|n\hspace{-1mm}-\hspace{-1mm}1\hspace{-0.25mm}], \hat{\phi}_{\mathrm{e}}[\hspace{-0.25mm}n|n\hspace{-1mm}-\hspace{-1mm}1\hspace{-0.25mm}]\hspace{-0.25mm}\right).\vspace{-1mm}
\end{equation}

\begin{figure}[t]
	\vspace{-5mm}
	\begin{minipage}{.49\textwidth}
		\center{\includegraphics[width=3.5in]{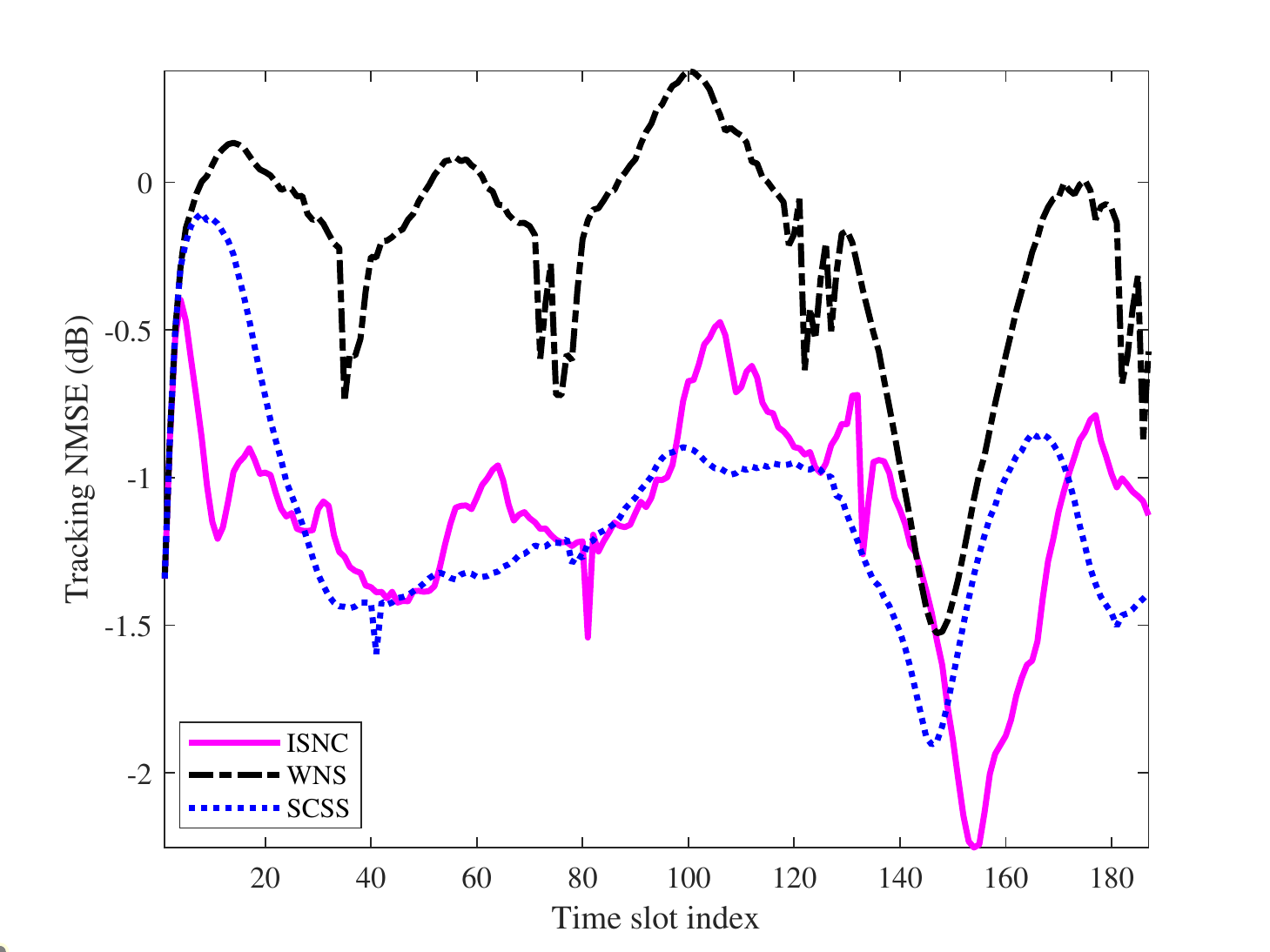}}\vspace{-7mm}
		\caption{Tracking NMSE of E-UAV state for the proposed ISNC scheme and the baseline schemes.}\vspace{-7mm}
		\label{fig:TrackingMSEMultiStep_Compare}	
	\end{minipage}	
	\hspace{1mm}
	\begin{minipage}{.49\textwidth}
		\center{\includegraphics[width=3.5in]{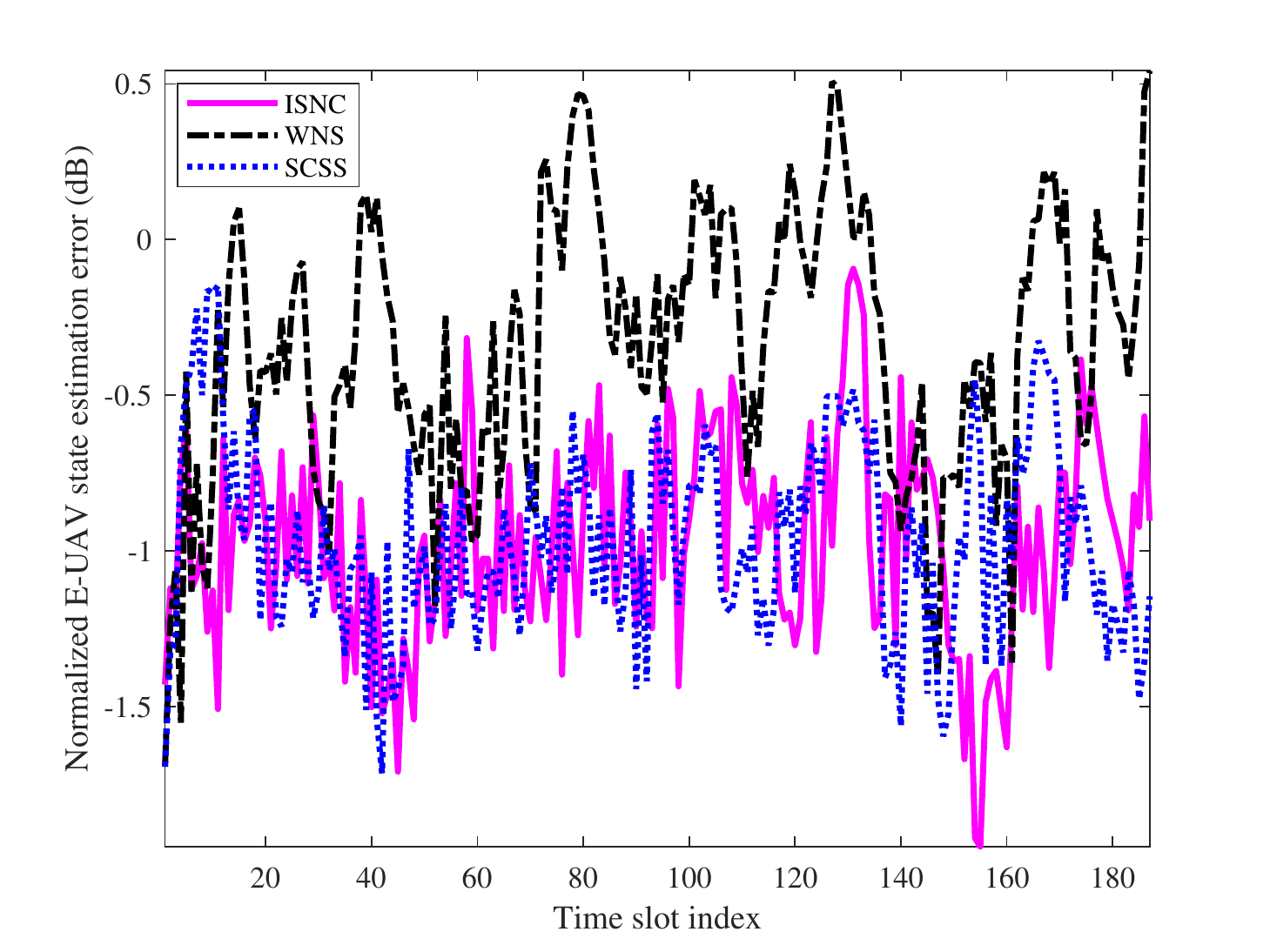}}\vspace{-7mm}
		\caption{E-UAV state estimation error for the proposed ISNC scheme and the baseline schemes.}\vspace{-7mm}
		\label{fig:EstimatedErrorMultiStep_Compare}	
	\end{minipage}
\end{figure}

\begin{figure}[t]
	\begin{minipage}{.49\textwidth}
		\center{\includegraphics[width=3.5in]{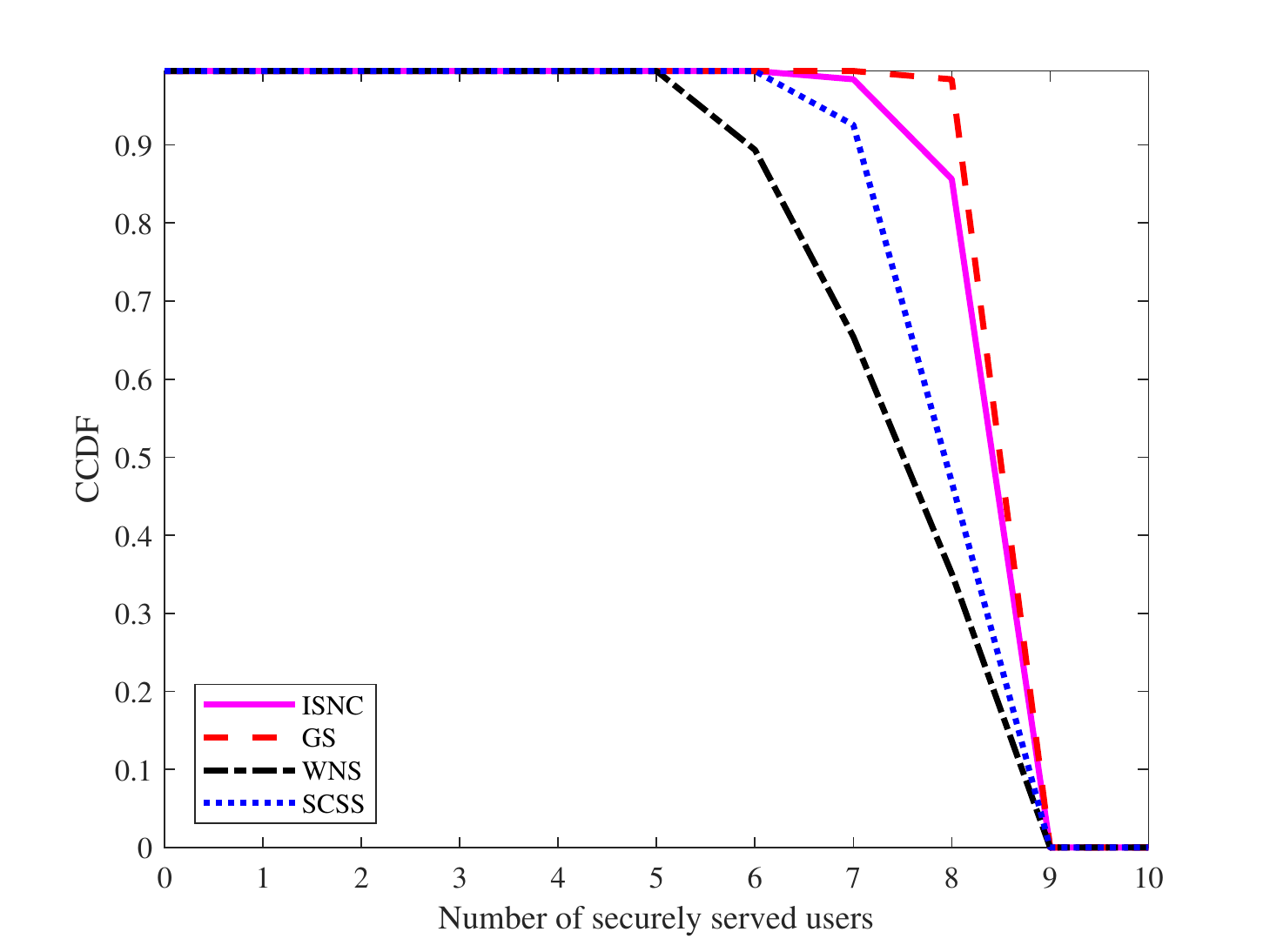}}\vspace{-7mm}
		\caption{The CCDF of the number of users securely served for the proposed ISNC scheme and the baseline schemes.}\vspace{-10mm}
		\label{fig:ComPerformanceMultiStep_Compare}	
	\end{minipage}
	\hspace{1mm}
	\begin{minipage}{.49\textwidth}
		\center{\includegraphics[width=3.5in]{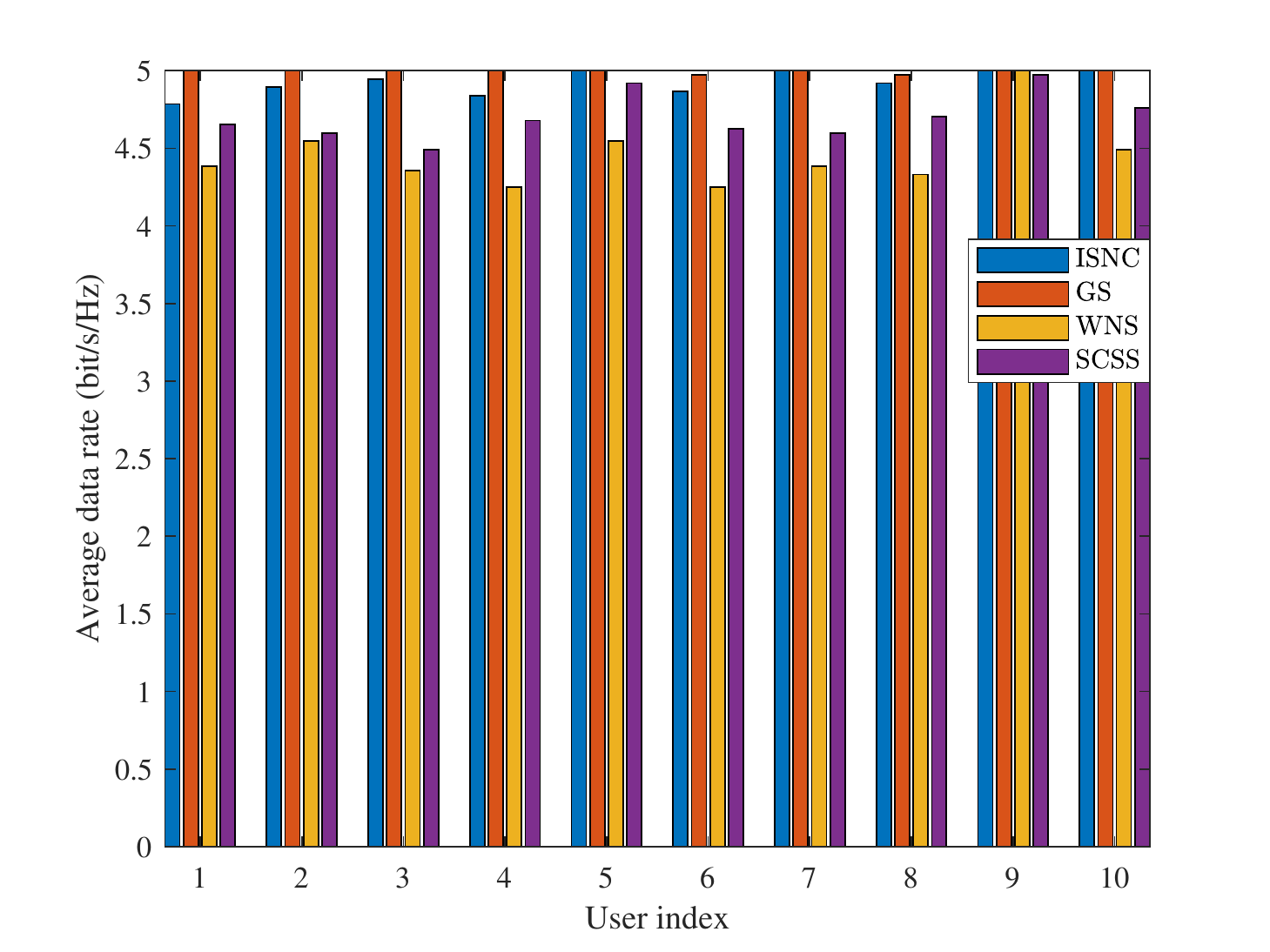}}\vspace{-7mm}
		\caption{Average data rate of each user for the proposed ISNC scheme and the baseline schemes.}\vspace{-10mm}
		\label{fig:RatePerformanceMultiStep_Compare}	
	\end{minipage}
\end{figure}

The designed trajectory of the I-UAV, the sensing performance, and the secure communication performance of the proposed ISNC scheme and the baseline schemes are provided in Figs. \ref{fig:NavigationandSensingBaseline}, \ref{fig:TrackingMSEMultiStep_Compare}, \ref{fig:EstimatedErrorMultiStep_Compare}, \ref{fig:ComPerformanceMultiStep_Compare}, and 
\ref{fig:RatePerformanceMultiStep_Compare}, respectively, with $L=5$ and $\lambda = 0.5$.
In Fig. \ref{fig:NavigationandSensingBaseline}, we can observe that the designed trajectories of the I-UAV for ``ISNC'' and ``SCSS'' are similar to that of ``GS'' as the navigation policies of ``ISNC'' and ``SCSS'' are mainly determined by the estimated trajectories of the E-UAV, which are well aligned with the ground-truth in both cases.
Besides, the sensing performance of ``WNS'' is much worse than that of ``ISNC'' and ``SCSS'', as can be seen in Figs. \ref{fig:TrackingMSEMultiStep_Compare} and \ref{fig:EstimatedErrorMultiStep_Compare}.
Allowing the I-UAV to move with a proper navigation design increases the MF output SNR and thus improves sensing performance.
Also, as illustrated in Figs. \ref{fig:TrackingMSEMultiStep_Compare} and \ref{fig:EstimatedErrorMultiStep_Compare}, the proposed ISNC scheme can achieve similar sensing performance as ``SCSS''. 
In contrast, as shown in Figs. \ref{fig:ComPerformanceMultiStep_Compare} and 
\ref{fig:RatePerformanceMultiStep_Compare}, the proposed ISNC scheme outperforms ``SCSS'' in terms of secure communication performance and approaches that of ``GS'' with a small gap.
This is due to the efficient exploitation of the system resources for simultaneous jamming and sensing by the proposed dual use of AN.
In addition, we can observe that ``WNS'' performs worst in terms of secure communication as it does not exploit the mobility of the I-UAV  for improving the performance of sensing, jamming, and communication.

\vspace{-4mm}
\subsection{Sensing and Communication Performance Trade-off}
\begin{figure}[t]
	\vspace{-5mm}
	\center{\includegraphics[width=4in]{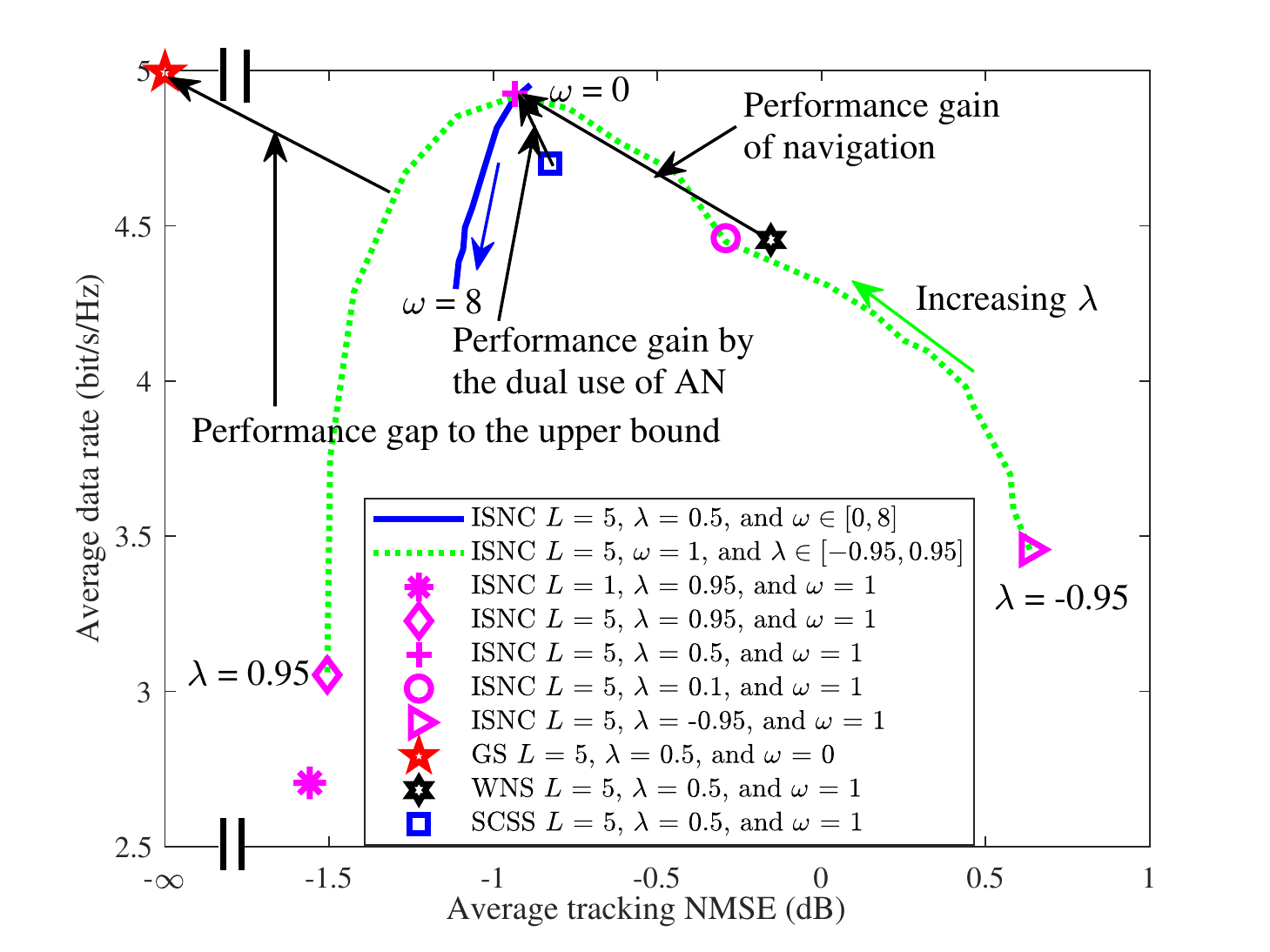}}\vspace{-7mm}
	\caption{The sensing and communication performance trade-off for the proposed ISNC scheme and baseline schemes.}\vspace{-10mm}
	\label{fig:Tradeoff}
\end{figure}

The sensing and communication performance trade-off in terms of average tracking NMSE and average data rate during all time slots is depicted in Fig. \ref{fig:Tradeoff}.
Note that in Fig. \ref{fig:Tradeoff}, the closer a scheme can approach the left-top corner, the better the sensing and communication performance trade-off, where the idealized ``GS'' can achieve the best trade-off.
As can be observed, for $\lambda = 0.5$,  increasing the resource allocation parameter $\omega$ reduces the average tracking NMSE slightly but degrades the average data rate significantly, since the resource allocation design in \eqref{Eqn:ProblemFormulation} places a higher priority on sensing.
Besides, for $\omega = 1$, increasing $\lambda$ from $-0.95$ to $0.95$  first benefits both sensing and communication, but then for $\lambda \ge 0.5$, further increasing $\lambda$ still reduces the tracking NMSE slightly but at the cost of a significantly degraded secure communication performance.
The achievable sensing and communication trade-off of all baseline schemes are also illustrated in Fig. \ref{fig:Tradeoff}.
We can observe that the proposed ISNC scheme can closely approach the optimal performance of ``GS'' with a small gap.
For instance, for $\lambda = 0.5$ and $\omega = 0$, the proposed ISNC scheme can achieve an average data rate of $4.95$ bit/s/Hz and an average tracking NMSE of $0.13$ (not in dB), which is close to the optimal performance of ``GS'' with an average data rate of $5$ bit/s/Hz and an average tracking NMSE of $0$ (not in dB).
Also, the performance of ``SCSS'' is strictly within the sensing and communication trade-off region achieved by the proposed ISNC scheme owing to the dual use of AN. 
Moreover, for $\lambda \le 0.5$, it can be seen that ``WNS'' performs worst which indicates that integrating navigation into the considered system is crucial and yields a significant gain in terms of both sensing and communication.

\vspace{-4mm}
\section{Conclusions}
In this paper, we proposed a novel ISNC framework for improving the secure communication performance of wireless UAV networks in the presence of a mobile E-UAV.
Through the dual use of AN, the state of the E-UAV was estimated and the wiretap channel was predicted.
The acquired knowledge was then exploited for online navigation and robust resource allocation design.
The proposed online navigation design minimized the distance between the I-UAV and a pre-defined desired destination point while taking into account kinematic and geometric constraints.
Based on the designed navigation policy of the I-UAV and the predicted location of the E-UAV, the wiretap channel between them was predicted and a fully-connected neural network was employed to determine a bound on the channel prediction error based on the E-UAV state estimation error.
Then, a robust resource allocation design was proposed to optimize the trade-off between sensing and communication in the next time slot considering the wiretap channel prediction error and the QoS requirements for secure communication.
Our simulation results confirmed the excellent performance of the proposed ISNC scheme compared to baseline schemes and provided some interesting insights.
In particular, (1) the dual use of AN is beneficial for improving both sensing and jamming; 2) integrating navigation into the considered system is crucial for improving both sensing and secure communication; 3) the navigation design has a larger impact on the trade-off between sensing and secure communication than the communication resource allocation design. 


\bibliographystyle{IEEEtran}
\bibliography{UAV_ISAV}

\end{document}